\documentclass{jpp}
\usepackage{graphicx}

\usepackage[utf8]{inputenc}
\usepackage[T1]{fontenc}
\usepackage{amsmath}
\usepackage{url}

\usepackage{booktabs, array, longtable, makecell}

\usepackage{pdflscape}

\shorttitle{Machine Learning Plasma Closure Review}
\shortauthor{S. Burles, E. Camporeale}

\title{The Machine Learning Approach to Moment Closure Relations for Plasma: A Review}

\author{Samuel Burles\aff{1}
  \corresp{\email{s.burles@qmul.ac.uk}},
  Enrico Camporeale\aff{1},\aff{2}}

\affiliation{\aff{1}School of Physical and Chemical Sciences, Queen Mary University of London,
London, E1 4NS, UK
\aff{2}Space Weather TREC, University of Colorado, Boulder, CO 80303, USA}

\begin{document}

\maketitle

\begin{abstract}
The requirement for large-scale global simulations of plasma is an ongoing challenge in both space and laboratory plasma physics. Any simulation based on a fluid model inherently requires a closure relation for the high order plasma moments. This review compiles and analyses the recent surge of machine learning approaches developing improved plasma closure models capable of capturing kinetic phenomena within plasma fluid models. We survey two methodological families: neural-network surrogates (from multilayer perceptrons to Fourier neural operators, the latter recently reproducing both linear and non-linear Landau damping online within a fluid solver) and equation-discovery methods such as sparse regression; and organise the studies by whether they are tested offline against reference data or online within a time-evolving solver. We outline the challenges associated with machine-learning closures, including off-diagonal pressure-tensor accuracy, generalisation beyond the training distribution, and stable integration into large-scale simulations, and the directions future research might take to address them.
\end{abstract}

\section{Introduction}

Space and laboratory plasmas exhibit complex dynamics across many spatio-temporal scales. For many phenomena observed to occur in plasmas, it is often the case that wave-particle interactions play a crucial role in the large-scale dynamics. Therefore, we must have accurate, multi-scale models to describe and predict these phenomena \citep{lapenta_we_2022, hammett_fluid_1990, lautenbach_multiphysics_2018}.

These dynamics span a hierarchy of characteristic scales. At the smallest scales, the Debye length $\lambda_D$ sets the limit of collective plasma behaviour. At intermediate scales, each species $\alpha$ is characterised by a gyroradius $\rho_\alpha = v_{t,\alpha}/\Omega_\alpha$ (where $v_{t,\alpha}$ is the thermal speed and $\Omega_\alpha = q_\alpha B / m_\alpha$ the cyclotron frequency) and an inertial length $d_\alpha = c / \omega_{p,\alpha}$ (where $\omega_{p,\alpha}$ is the plasma frequency). The large mass ratio between ions and electrons produces a separation between scales for each species: the electron gyroradius and inertial length ($\rho_e$, $d_e$) are much smaller than their ion counterparts ($\rho_i$, $d_i$), which are in turn much smaller than the system scale $L_\mathrm{sys}$. Each regime therefore demands a different level of physical fidelity from the plasma model, a theme that recurs throughout this review.

Historically, the numerical modelling of plasma has been broadly divided between kinetic and fluid approaches. Kinetic models of collisionless plasmas, typically found in space environments, are described by the time evolution of a distribution function, governed by the collisionless Vlasov equation. The so-called one-particle distribution function describes the probability distribution of particles in velocity-position phase space. Fully kinetic simulations can capture self-consistently the interactions between particles and electromagnetic fields. However, they are computationally prohibitive at large scales because of the dimensionality of the phase space, and the stringent requirements on time and grid resolution, for numerical stability.

A fluid description of plasma consists of the velocity moments of the distribution function, which describe the plasma by a series of observable macroscopic quantities. Fluid descriptions of plasmas are low-dimensional and therefore better suited to model large-scale systems while maintaining a low computational cost. Fluid simulations, however, are ill-suited to resolve kinetic effects due to the simplifying approximations required to construct the models. The accuracy of a fluid description depends on how well the effects of higher-order moments, which are not explicitly retained, can be captured through an appropriate closure. Fluid models tend to perform better when the underlying distribution is close to Maxwellian, but this is a consequence of the closure approximations rather than an explicit assumption of the framework itself.

Plasma fluid models can be rigorously derived by taking velocity moments of the Vlasov equation, resulting in an infinite hierarchy of equations, where the evolution of each moment is dependent on the next higher-order moment, leaving the set of equations open. Hence, to obtain a self-consistent fluid model, one must truncate the system of equations by specifying a closure. A closure equation specifies a chosen highest-order moment in terms of only lower-order moments. The kinetic effects will typically be lost when the physical assumptions necessary to obtain an analytic closure model are made. It is therefore a key problem to find new closure models that can include important kinetic effects in fluid models without incurring a steep computational cost in simulations.

One aspect of plasma modelling that has always posed a computational challenge is the different scales associated with various species in the plasma. Electrons evolve on much shorter time scales than ions owing to the significant difference in the charge-to-mass ratio. To model both ion and electron species, a hybrid model can be employed where one species is treated kinetically and one as a fluid, with the ion species often chosen to be kinetic and the electron species as a fluid. Another option is to lower the simulated difference in charge-mass ratio between species to speed up the computation \citep{buchner_hybrid_2003, taccogna_hybrid_2016, palmroth_magnetosheath_2018, palmroth_vlasov_2018, valentini_hybrid-vlasov_2007}.

A further strategy for modelling large-scale systems while partly accounting for kinetic effects is to embed a small kinetic domain into a larger fluid simulation. For example, in the context of a planetary magnetosphere, the magnetotail is typically treated kinetically because it is the region where magnetic reconnection is expected to occur \citep{wang_global_2022, toth_extended_2016, lautenbach_multiphysics_2018, keebler_simulating_2025, wang_electron_2024, chen_fleks_2023}.

\begin{figure}
  \centering
  \includegraphics[width=0.6\textwidth]{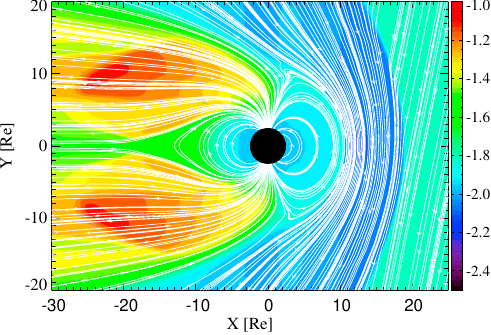}
  \caption{Spatial distribution of the ion inertial length~$d_i$ in a global magnetosphere simulation, illustrating the large variation in kinetic scales across the simulation domain. Regions where~$d_i$ is comparable to the local grid resolution require kinetic treatment, motivating the embedded kinetic-fluid simulation approach. Reproduced from \citet{toth_extended_2016}, \textit{J.\ Geophys.\ Res.\ Space Phys.} \textbf{122}, 10,336-10,363, Fig.~2.}
  \label{fig:toth_inertial_length}
\end{figure}

Recent advances in parallel computing and GPU capabilities have driven rapid progress in machine learning, resulting in new techniques applied in multiple scientific fields. Machine learning approaches are becoming increasingly popular in neutral fluid dynamics, with much focus on improving turbulent simulations and the long-lasting closure problem \citep{brunton_machine_2020, sanderseScientificMachineLearningSatMar0100:00:00UTC2025, li_graph_2022, vinuesa_enhancing_2022}.

This review paper aims to collect and analyse various works in scientific machine learning and plasma simulation that focus on developing improved closure models that include kinetic physics. The review will begin with a brief overview of the background physics involved in defining the plasma moment closure. In Section 2, an overview of the relevant machine learning techniques and methodology will be given to provide context. Section 3 presents a review of the research that has been conducted in applying a data-driven approach to solving the plasma closure problem. Finally, Section 4 concludes and presents future ideas.

\subsection{The Closure Problem}

In the kinetic description of plasma, one can, in principle, describe the system of particles with a function containing all particle positions and velocities at a given time, $N_{\alpha}(\boldsymbol{x}, \boldsymbol{v}, t)$. This detailed description of a plasma, while containing all necessary information, is, however, difficult to work with, and so we seek a reduced description of the plasma. This is achieved by performing a statistical average over particles. We restrict our model to the one-particle distribution function, neglecting particle interactions. The one-particle distribution function $f_{\alpha}(\boldsymbol{x}, \boldsymbol{v}, t)$ describes the probability density that a particle of species $\alpha$ will have a position $\boldsymbol{x}$ and velocity $\boldsymbol{v}$ within an infinitesimal region of phase-space $d\boldsymbol{x}d\boldsymbol{v}$, at time $t$.

To describe the plasma dynamics, we then require an evolution equation of the distribution function, self-consistently coupled with the Maxwell equations. This evolution equation is the Vlasov equation and can be derived through the so-called BBGKY procedure, starting from the Liouville equation for the one-particle distribution function, or from taking statistical averages of the Klimontovich equation for the density function \citep{krall_principles_1973, swanson_plasma_2003, fitzpatrick_plasma_2014}. We further restrict ourselves to the non-relativistic collisionless Vlasov equation \citep{vlasov_vibrational_1968}:
\begin{equation}\label{eq:Vlasov}
    \frac{\partial f_{\alpha}}{\partial t} + \boldsymbol{v} \bcdot \bnabla f_{\alpha} + \frac{q_{\alpha}}{m_{\alpha}} \left(\boldsymbol{E} + \boldsymbol{v} \times \boldsymbol{B}\right) \bcdot \frac{\partial f_{\alpha}}{\partial \boldsymbol{v}} = 0
\end{equation}
coupled to the Maxwell equations:
\begin{gather}
    \bnabla \bcdot \boldsymbol{E} = \frac{1}{\epsilon_{0}} \sum_{\alpha} q_{\alpha} \int d \boldsymbol{v} f_{\alpha}(\boldsymbol{x}, \boldsymbol{v}, t)\\
    \bnabla \bcdot \boldsymbol{B} = 0\\
    \bnabla \times \boldsymbol{E} + \frac{\partial \boldsymbol{B}}{\partial t} = 0\\
    \bnabla \times \boldsymbol{B} - \mu_{0} \epsilon_{0} \frac{\partial \boldsymbol{E}}{\partial t} = \mu_{0} \sum_{\alpha} q_{\alpha} \int d \boldsymbol{v} \; \boldsymbol{v} f_{\alpha}(\boldsymbol{x}, \boldsymbol{v}, t)
\end{gather}
where $\boldsymbol{E}$ and $\boldsymbol{B}$ are the electric and magnetic fields, respectively, $q_{\alpha}$ and $m_{\alpha}$ are the charge and mass of a particle of species $\alpha$, $\epsilon_0$ is the electric permittivity of the vacuum and $\mu_0$ is the magnetic permeability of the vacuum.

The kinetic description of a collisionless plasma, therefore, requires a 6-dimensional phase space, plus time in order to be simulated. This high dimensionality often makes the computational cost of large-scale 3D simulations using the kinetic description prohibitive, even on the most advanced HPC architecture.

Techniques such as the particle-in-cell (PIC) method for computing fully kinetic simulations can mitigate some of the computational cost by utilising 'macroparticles' to discretise the distribution function. PIC methods reduce computational cost compared to Eulerian Vlasov codes by representing the distribution function with a finite number of macroparticles rather than discretising the full six-dimensional phase space on a grid, thereby circumventing the exponential growth of grid points with dimensionality. PIC methods do still, however, suffer from the high-dimensional phase space, and are not yet suited for long-time, global simulations \citep{markidis_multi-scale_2010, germaschewski_plasma_2016}.

To reduce the computational cost required to simulate a plasma, it makes sense to seek reduced models. As we mentioned, fluid equations can be constructed by taking velocity moments of the particle distribution function.

Following the procedure set out in \citet{swanson_plasma_2003}, we can write the collisionless Vlasov equation as
\begin{equation}
    \frac{\partial f}{\partial t} + \bnabla \bcdot (\boldsymbol{v}f) + \bnabla_{\boldsymbol{v}} \bcdot (\boldsymbol{A}f) = 0,
\end{equation}
where $\boldsymbol{A}$ is the acceleration due to the electric and magnetic fields, and the subscripts $\alpha$ have been dropped for conciseness. We can proceed to introduce a scalar function of velocity, $Q_{m}(\boldsymbol{v}) = \boldsymbol{v}^{m}$. We define the moment taking process as averaging over the velocity:
\begin{equation}
    \left<Q_{m}(\boldsymbol{v})\right> \; = \; \int Q_{m} f d\boldsymbol{v}.
    \label{average_def}
\end{equation}
We can then define the zeroth order moment $n(\boldsymbol{x}) = \int f(\boldsymbol{x},\boldsymbol{v})d\boldsymbol{v}$ as the density in configuration space, corresponding to $\left< Q_0 \right>$. We can then proceed by multiplying the Vlasov equation by $Q_{m}(\boldsymbol{v})$ and integrate in velocity:
\begin{equation}
    \int Q_{m} \frac{\partial f}{\partial t} d\boldsymbol{v}+ \int Q_{m} \bnabla \bcdot \boldsymbol{v} f d\boldsymbol{v} + \int Q_{m} \bnabla_{\boldsymbol{v}} \bcdot \boldsymbol{A} f d \boldsymbol{v} = 0.
\end{equation}

As $Q_{m}(\boldsymbol{v})$ is a function only of the velocity, we can rearrange as:
\begin{equation}
    \frac{\partial}{\partial t} \int Q_{m} f d\boldsymbol{v} + \bnabla \bcdot \int Q_{m} \boldsymbol{v} f d\boldsymbol{v} + \int Q_{m} \bnabla_{\boldsymbol{v}} \bcdot \boldsymbol{A} f d\boldsymbol{v} = 0.
\end{equation}

It is then clear that the first term is simply $\frac{\partial}{\partial t} \left<Q_{m}\right>$ from the definition in Eq. (\ref{average_def}). The second term can be written similarly as $\bnabla \bcdot \left<Q_{m}\boldsymbol{v}\right>$ as $Q_{m}$ is a scalar function of velocity only. The third term can be simplified by using the vector identity $\bnabla \bcdot (a\boldsymbol{A}) = \boldsymbol{A} \bcdot \bnabla a + a \bnabla \bcdot \boldsymbol{A}$ and the divergence theorem. Then, assuming that the distribution vanishes for $v \rightarrow \infty$, causing the surface integral to vanish, the third term becomes $\left< \boldsymbol{A} \bcdot \bnabla_{\boldsymbol{v}} Q_{m} \right>$. Writing all terms together gives:
\begin{equation}
    \frac{\partial}{\partial t} \left<Q_{m}\right> + \bnabla \bcdot \left< Q_{m} \boldsymbol{v} \right> - \left< \boldsymbol{A} \bcdot \bnabla_{\boldsymbol{v}} Q_{m} \right> = 0.
\end{equation}

This is a general formula for the evolution of the velocity moments. By setting the scalar function $Q_{m}(\boldsymbol{v})$ to different powers of the velocity, we can obtain the evolution equation for the moments of arbitrary order. The closure problem arises because the second term of this general equation will always contain the next-higher order moment in $\left<Q_{m}\boldsymbol{v}\right>$. The hierarchy of equations can only be closed by defining the last-order moment as a function of only lower-order moments: that constitutes our closure. Although the concept of closure is relatively straightforward to state, different choices of closure dramatically affect the accuracy of a model and how well (or not) kinetic effects are captured. In addition to this, the derivation of analytic closure relations can become very non-trivial, causing the search for plasma closure relations to remain open.

Finally, it is customary to define macroscopic fluid quantities in terms of the first four velocity moments \citep{wang_comparison_2015, donaghy_search_2023}:
\begin{align}
    n &= \int f d\boldsymbol{v}, \\
    \boldsymbol{u} &= \frac{1}{n} \int \boldsymbol{v} f d\boldsymbol{v}, \\
    \mathsfbi{P} &= m \int (\boldsymbol{v} - \boldsymbol{u})(\boldsymbol{v} - \boldsymbol{u}) f d\boldsymbol{v}, \\
    \mathsfbi{Q} &= m \int (\boldsymbol{v} - \boldsymbol{u}) (\boldsymbol{v} - \boldsymbol{u}) (\boldsymbol{v} - \boldsymbol{u}) f d\boldsymbol{v}.
\end{align}

The zeroth-order moment is the bulk density, the first-order moment is the bulk velocity, the second-order moment is the pressure tensor, and the third-order moment is the heat flux tensor. $\mathsfbi{P}$ and $\mathsfbi{Q}$ have been defined as the centred-difference moments of the distribution function to define the observable fluid properties.

The heat flux in particular plays a central role in regulating energy conversion in weakly collisional and collisionless plasmas \citep{svenningssonElectronHeatFlux2026a}, making its faithful representation a key motivation for the improved closures discussed below.

The fluid model for the first three moments becomes \citep{hunana_introductory_2019}:
\begin{gather}
    \frac{\partial n}{\partial t} + \bnabla \bcdot (n \boldsymbol{u}) = 0, \\
    m \left[ \frac{\partial \boldsymbol{u}}{\partial t} + (\boldsymbol{u} \bcdot \bnabla) \boldsymbol{u} \right] + \left( \frac{1}{n} \right) \bnabla \bcdot \mathsfbi{P} - q(\boldsymbol{E} + \boldsymbol{u} \times \boldsymbol{B}) = 0, \\
    \frac{\partial \mathsfbi{P}}{\partial t} + \bnabla \bcdot (\boldsymbol{u}\mathsfbi{P} + \mathsfbi{Q}) + \mathsfbi{P} \bcdot \bnabla \boldsymbol{u} + (\mathsfbi{P} \bcdot \bnabla \boldsymbol{u})^{\text{T}} + \frac{q}{m}[\boldsymbol{B} \times \mathsfbi{P} + (\boldsymbol{B} \times \mathsfbi{P})^{\text{T}}] = 0.
\end{gather}
Once again, it is evident how the evolution of each moment depends on the next higher-order moment.

Many examples of closure exist, with each typically intended to model a different plasma regime or phenomena. A very comprehensive and detailed review from \citet{hunana_introductory_2019, hunana_introductory_2019-2} is recommended for readers interested in the details of the derivation of the various models. We include here a more brief review of analytical plasma closures, to provide the context for which we investigate the use of machine learning methods.

\subsection{A Brief History of Plasma Closures}

The traditional approach to the plasma closure relation is finding analytical closure equations, typically by deriving relations from the kinetic theory and employing an expansion of the distribution function in order to close the system of equations. Analytical plasma closure relations have been developed from simple adiabatic fluid models into increasingly sophisticated frameworks that embed key kinetic effects such as pressure anisotropy, finite Larmor radius (FLR) physics, and Landau damping within tractable moment hierarchies. Each stage in this development has extended the range of plasma regimes accessible to fluid models. In this way, however, fluid models tend to be relatively \textit{ad hoc}, and will generally be applicable only to the phenomena or regime the closure was designed for.

As previously stated, the past work on the closure problem has resulted in plasma fluid models that can roughly be grouped into different classes. This review will focus on the adiabatic closure models, Landau fluid models and gyrofluid models.

A given closure can be applied to different fluid models, based on which moment in the hierarchy the closure is specified, and on the derivation of the moment evolution equations. Commonly, fluid models are referenced by the number of moments included from the hierarchy. For example, a 5-moment fluid model denotes a fluid model including the 0th order moment, $n$, the 3 components of the 1st order moment, $(u_x, u_y, u_z)$, and a scalar for the 2nd order moment, $p$, giving 5 total moments. The closure can then take any form, describing the scalar pressure $p$. A 10-moment model includes the full symmetric pressure tensor, thus describing 10 different moment components.

As the discussion thus far has been concerned with the development of reduced plasma models for the purpose of reducing the computational load of simulations without sacrificing physical fidelity, it is worth noting the Darwin approximation. The computational cost of plasma simulations is strongly influenced by the electromagnetic formulation employed. In particular, the full Maxwell equations allow for the propagation of electromagnetic waves at the speed of light, $c$, which imposes a stringent constraint on the simulation timestep through the Courant-Friedrichs-Lewy (CFL) condition. This limitation is dictated by numerical instability and often leads to excessively small timesteps, substantially increasing computational expense. The Darwin approximation offers a practical alternative by neglecting the transverse component of the displacement current in Amp\`{e}re’s law. This modification effectively removes the propagation of transverse electromagnetic (light) waves while retaining the essential quasi-static magnetic and electrostatic effects relevant to sub-relativistic plasma dynamics. As a result, the CFL constraint is relaxed, allowing significantly larger timesteps and thus reducing the overall computational load without compromising the fidelity of low-frequency plasma processes. Crucially, any 10-moment (or higher) plasma fluid model will include transverse components of the moment tensors, leading to transverse electromagnetic waves, and so will be affected by the CFL condition for the propagation of these waves at speed $c$. A Darwin approximation will also need to be considered for these plasma fluid models \citep{pezzi_vida_2019, kaufman_darwin_1971, mangeney_numerical_2002, schmitz_darwinvlasov_2006}.

\subsubsection{Adiabatic Fluid Models}

The earliest work on collisionless plasma closure relations came from the paper of Chew-Goldberger-Low \citep{chew_boltzmann_1956} in adding anisotropic pressure to plasma fluid models. The CGL closure is derived from an expansion of the distribution function in powers of the mass-charge ratio of ions (equivalent to an expansion in powers of the Larmor radius) and only the leading order is taken. Only ions are retained, with the electrons being considered as a macroscopic neutralising background. The CGL model is a plasma fluid which is able to capture pressure anisotropy. The pressure is split into parallel and perpendicular components with respect to the magnetic field, $p_{\parallel}$ and $p_{\perp}$ respectively. The CGL model closes the moment equations at the second order, setting the heat-flux to zero, resulting in the conservation of two adiabatic invariants, the longitudinal action, and the magnetic moment.
\begin{align}
    \frac{\mathrm{d}}{\mathrm{d}t} \left(\frac{p_{\parallel}B^2}{\rho_0^3} \right) = 0, \\
    \frac{\mathrm{d}}{\mathrm{d}t} \left(\frac{p_{\perp}}{\rho_0 B} \right) = 0.
\end{align}
Here $\rho_0$ is the mass density calculated from the 0th order distribution function in the mass-charge expansion, and $B$ is the magnitude of the magnetic field. The class of plasma fluid models known as adiabatic models are extensions of the CGL model.

Physically, the CGL double-adiabatic invariants partly account for the pressure anisotropies observed in the expanding solar wind; however, as anisotropy grows it raises the free energy available to kinetic micro-instabilities (e.g. mirror and firehose), which act to scatter particles and return the plasma toward isotropy; purely adiabatic closures cannot capture this feedback.

Early work based on the CGL model was performed by \citet{abraham-shrauner_propagation_1967} in extending the CGL model to including a polytropic equation of state for the parallel and perpendicular pressures. The extended model was then used to investigate the difference in phase speeds of plasma waves between magnetohydrodynamics (MHD) and CGL models. \citet{abrahamshrauner_doubleadiabatic_1968} argued that the electrons must be included explicitly with their fluid moments, rather than only as a macroscopic current, as the electric field parallel to the magnetic field frequently cannot be neglected, as it was in \citet{chew_boltzmann_1956}. \citet{abraham-shrauner_small_1973} then introduced an extension of the CGL model by introducing a generalized polytropic law for anisotropic pressure components utilising 4 polytropic indices. The polytropic approach allows the model to be tested in regimes ranging MHD-like isotropic pressure, to the anisotropic pressure of CGL:
\begin{align}
    p_{\perp} &= K_{\perp} \rho^{\epsilon} B^{\gamma}, \\
    p_{\parallel} &= \frac{K_{\parallel} \rho^{\beta}}{B^{\alpha}},
\end{align}
where $K_{\perp}$, $K_{\parallel}$, $\alpha$, $\beta$, $\gamma$ and $\epsilon$ are constants.

A direct application followed in \citet{ballai_linear_2002}, who used a simpler two-index polytrope law to derive dispersion relations for both linear and non-linear waves in a slab. Dispersion relations were derived for both the linear and non-linear waves present in the adiabatic fluid model, investigating the specific case of plasma waves in a slab.

\citet{le_equations_2009} departed from the models discussed so far, deriving a closure from an analysis of electron trapping in order to capture magnetic reconnection in a fluid model. Beginning from a gyroaveraged distribution function (averaged over the fast gyrophase of the cyclotron motion) and the parallel electric potential the authors gave the distribution in the reconnection region for both trapped and transient particles as of that in the ambient plasma via Liouville's theorem. \citet{le_equations_2009} then gave an expansion of the distribution function about a leading order Maxwellian, in order to formulate a fluid description where any order moment can be recast as a function of $B$ and $n$, the magnetic field magnitude and 0th order moment respectively.
\begin{equation}
    \frac{n}{n_{\infty}} = \frac{2}{\sqrt{\pi}} (1-b)\sqrt{u} + e^{u} \Psi(\sqrt{u}) - b^{3/2} e^{u/b} \Psi \left(\sqrt{\frac{u}{b}} \right), \label{le_2009}
\end{equation}
where $u=e\Phi_{\parallel}/T_{\infty} > 0$ and $b = 1 - B/B_{\infty} > 0$. $\Psi$ is the complementary error function. $X_{\infty}$ denotes the value of the variable in the ambient plasma and $T_{\infty}$ is the ambient plasma temperature. Inverting equation \eqref{le_2009} numerically gives $\Phi_{\parallel}(n, B)$, which can then be used in the distribution functions for trapped and passing particles, from which higher-order moments can be calculated numerically. For trapped electrons, \citet{le_equations_2009} recovered the adiabatic invariants of the CGL model for negligible heat flux. Within this framework the parallel and perpendicular pressure components can be written in the approximate normalised form
\begin{equation}
  p_{*\parallel} = n_* \frac{2}{2+\alpha} + \frac{\pi^3 n_*^3}{6 B_*^2}\,\frac{2\alpha}{2\alpha+1},
  \label{eq:le_ppar}
\end{equation}
\begin{equation}
  p_{*\perp} = n_* \frac{1}{1+\alpha} + n_* B_*\,\frac{\alpha}{\alpha+1}
  \label{eq:le_pperp},
\end{equation}
where $\alpha = n_*^3/B_*^2$ and all quantities $X$ are normalised as $X_* = X/X_\infty$, with $X_\infty$ denoting the upstream value.

The \citet{le_equations_2009} closure was tested by \citet{ohia_demonstration_2012}, who compared a self-consistent fluid simulation against a fully kinetic one. They performed a self-consistent fluid simulation, with a simple isotropic closure for the ions and the anisotropic \citet{le_equations_2009} closure for electrons, and compared to a fully kinetic simulation focusing on magnetic reconnection. An isotropic pressure model was also compared as a baseline. Good agreement was found between the anisotropic fluid closure and the fully kinetic simulation. In doing so, they showed a closure intended for collisionless magnetic reconnection working well in a self-consistent fluid model, allowing for less computationally intense simulation as well as the ability to elucidate the physical processes in the reconnection region more easily than with the fully kinetic model.

A hybrid-code implementation of the anisotropic electron-pressure closure was provided by \citet{le_hybrid_2016}, with kinetic ions and fluid electrons. They used this hybrid model to simulate magnetic reconnection from both a force-free current sheet and a Harris sheet. They found that the anisotropic model gave a closer match than the isotropic electron-pressure closure and allows the formation of an extended current sheet in the reconnection region, as observed in the fully kinetic simulations.

\citet{wetherton_validation_2019} offered an observational test, validating the \citet{le_equations_2009} closure for guide-field reconnection against \textit{in situ} Magnetospheric Multiscale (MMS) data. \citet{wetherton_validation_2019} found that, for a single guide-field reconnection event, the observations agree well with the \citet{le_equations_2009} equation of state. The event is a current-sheet crossing lasting a few seconds, observed by MMS on 31 October 2015 within a magnetopause magnetic flux rope where two reconnection jets collide in a compressed current sheet under a strong guide field of roughly twice the reconnecting field \citep{oierosetMMSObservationsLarge2016}; the agreement therefore validates the closure for one specific reconnection current sheet rather than establishing a general closure for the magnetosheath or other regimes. The MMS data was also compared to simulations, both a hybrid simulation using the \citet{le_equations_2009} and a fully kinetic simulation. \citet{wetherton_validation_2019} found that the hybrid simulation matched the MMS data for this event more than the fully kinetic simulation, leading to the conclusion that the massless electron assumption of the \citet{le_equations_2009} closure was able to more closely match the observed dynamics than the fully kinetic simulation, which uses an ion-electron mass ratio of 100 for computational cost, potentially causing the discrepancy. Utilising the closure, \citet{wetherton_validation_2019} were also able to infer the properties of the upstream plasma from the MMS observations.

The adiabatic invariants provide useful physical insight into collisionless plasmas, at the cost of requiring a negligible heat flux and thus a truncation of the system of moment equations at the pressure, neglecting all higher-order moments. The adiabatic models also fail to capture any nongyrotropic behaviour (i.e.\ departures from rotational symmetry of the pressure tensor about the local magnetic field). The neglect of higher order terms in the Larmor radius expansion of the distribution function also prevents the models from capturing dynamics below the ion gyroradius. The addition of these smaller scales to the models is known as Finite Larmor Radius (FLR) Corrections.

The first FLR correction to the CGL model came from \citet{yajima_effect_1966}, who derived the dispersion relations and equations for small-amplitude oscillations. This was done by including the first-order term in the expansion of the distribution function in the Larmor radius (where only 0th order is taken in the CGL description) and linearizing the equations. \citet{yajima_effect_1966} found that the dispersion relation for small amplitude waves is dependent on both an FLR correction term and an electron inertia term, finding that within certain limits, the small amplitude wave equation behaves analogously to the Schrödinger equation.

A more recent FLR extension of an adiabatic model is that of \citet{hazeltine_local_2013}, who derived an expression for the internal energy of a magnetised, anisotropic plasma via a thermodynamic approach. They obtained an equation of state containing identifiable contributions from the CGL model (pressure anisotropy), gyroviscous tensor (collisional viscosity) and FLR contributions.

\subsubsection{Landau Fluid Models}

A key property of kinetic plasma is the process of phase mixing, leading to the collisionless damping of oscillations known as Landau damping. Prior to the work of \citet{hammett_fluid_1990}, it was assumed that capturing the kinetic processes of phase mixing and collisionless damping required fully-kinetic simulations. Fluid equations were believed fundamentally incapable of representing these processes. Hammett and Perkins demonstrated that an improved closure approximation could remedy this limitation.

\citet{hammett_fluid_1990} presented an electrostatic Landau-fluid closure derived from a multipole Pad\'{e} expansion of the linear plasma response function. This approach results in a non-local closure, most easily described in Fourier space:
\begin{equation}
    \tilde{q}_{k} = - n_{0} \sqrt{\frac{8}{\upi}} v_{t} \frac{\mathrm{i} k \tilde{T}_{k}}{|k|},
    \label{hp-closure}
\end{equation}
where the closure is of the 3rd order moment, the heat flux perturbation, $\tilde{q}$, $n_{0}$ is the equilibrium density, $v_{t}$ the thermal velocity of the particle species, $k$ the wavenumber and $\tilde{T}$ the temperature perturbation. Here, tilde denotes Fourier transformed quantities defined from the linear kinetic response. The $1/|k|$ dependence causes the closure to be non-local in configuration space. This non-locality renders the numerical evaluation of the closure computationally demanding, due to requiring a full integration over configuration space (prohibiting parallelisability of the fluid code for GPU acceleration) or, if computing in Fourier space, both a forward and inverse Fourier transform. Even using the fast Fourier transform (FFT) this will still cause a cost of $\mathcal{O}(N\log(N))$, compared to a local closure cost $\mathcal{O}(N)$. This increased cost of the non-local closure has led to the development of some local approximations that replace the $1/|k|$ dependence of the HP closure with a local approximation, typically replacing the wavenumber with a characteristic wavenumber $k_0$, where $1/k_0$ represents the length over which collisionless damping would be expected to occur.

A further difficulty arises for field-aligned non-local closures in geometries with chaotic magnetic field lines, where the parallel coordinate along which the non-local integral must be evaluated is itself stochastic. In such regions, the closure integral may become numerically ill-defined, as the integration path can diverge exponentially, posing a fundamental challenge beyond the computational cost considerations discussed above.

In a companion paper, \citet{hammett_fluid_1992} gave a fuller account of how to construct fluid models capable of Landau damping using the non-local closure of \citet{hammett_fluid_1990}. \citet{hammett_fluid_1992} demonstrated how introducing phase mixing at different orders of the velocity moments in the hierarchy affects the ability of the fluid models to approach kinetic models, with higher-order moment closures leading to models that more closely approach the kinetic results.

Building on Kulsrud's collisionless-MHD approach, \citet{snyder_landau_1997} expanded the Vlasov-Maxwell equation in small gyroradius before applying a guiding-centre treatment to obtain a set of fluid equations. This yielded the first Landau-fluid closure for magnetised plasma, concentrating on MHD scales. They took the CGL-like fluid model and instead implemented a HP-like Landau fluid closure for the heat flux (the `$3+1$' moment model) or the next-highest-order moment ($r$, described as the `$4+2$' moment model) by taking a two-pole Pad\'{e} approximation of the response function, similarly to \citet{hammett_fluid_1990}. The `3+1' moment closure is as follows for the parallel and perpendicular heat fluxes in Fourier space:
\begin{align}
    q_{\parallel} &=  -n_0 \sqrt{\frac{8}{\pi}} v_{t_{\parallel}} \frac{\text{i} k_{\parallel} T_{\parallel}}{|k_{\parallel}|}, \\
    q_{\perp} &= -n_0 \sqrt{\frac{2}{\pi}} v_{t_{\parallel}} \frac{\text{i} k_{\parallel} T_{\perp}}{|k_{\parallel}|} + n_0 \sqrt{\frac{2}{\pi}} v_{t_{\parallel}} T_{\perp_0} \left( 1 - \frac{T_{\perp_0}}{T_{\parallel_0}} \right) \frac{\text{i} k_{\parallel} B_1}{|k_{\parallel}| B_0},
\end{align}
where 0 subscripts refer to the zeroth-order distribution (defined as a bi-Maxwellian) and 1 subscripts to the first-order perturbation from the leading order distribution. Subscripts for the moment perturbations have been omitted for clarity. $B$ is the magnitude of the magnetic field, $T$ the temperature, $n$ the particle density, $k$ the wavenumber and $v_t$ the thermal velocity. The proportionality of $\text{i}k_\parallel/|k_\parallel|$ can be defined as the negative Hilbert transform operator $\mathcal{H}$. The \citet{snyder_landau_1997} closures are significant as one of the earliest examples of a Landau fluid model for magnetised plasma. \citet{snyder_landau_1997} applied their closure to reconstruct the mirror instability threshold and reproduce the MHD-wave dispersion relations for a collisionless bi-Maxwellian distribution.

As mentioned, the non-local nature of the HP closure causes models to be computationally prohibitive, due to the need to integrate over all physical space, or along field lines. In an effort to capture the important collisionless damping with a less computationally demanding closure, \citet{wang_comparison_2015} developed both a 5- and 10- moment 2-fluid model using a local approximation of the HP closure approach extended to magnetised plasma. In the local approximation, the wavenumber dependence of the HP closure (leading to the non-local behaviour) is replaced by a chosen characteristic wavenumber, representing lengths over which collisionless damping is expected to occur. The 5-moment fluid model given by \citet{wang_comparison_2015} closes the equations by setting the heat-flux and agyrotropic pressure tensors to 0, resulting in a scalar pressure. The 10-moment fluid model is defined from quantities perturbed about equilibrium quantities, with the closure of the heat flux tensor perturbation taking the form in physical space:
\begin{equation}
    \partial_m Q_{ijm} \approx v_t |k_0| (P_{ij} - p\delta_{ij}),
\end{equation}
where $Q_{ijm}$ is the components of the heat flux tensor, $v_t$ is the thermal velocity of the species, $k_0$ is the characteristic wavenumber chosen in the local approximation, $P_{ij}$ is the full pressure tensor and $p=P_{ii}/3$ is the scalar pressure. \citet{wang_comparison_2015} compared results of simulations using these local models to that of a PIC simulation of collisionless reconnection of a Harris sheet, as a means of testing the effectiveness of the local approximation. It was demonstrated that a 10-moment model, even with only a local approximation for the closure, was able to predict the dynamics of the reconnection region consistent with that of the kinetic PIC simulation.

The \citet{wang_comparison_2015} local closure was deployed at global scale by \citet{dong_global_2019}, in a multi-species ten-moment model of Mercury's magnetosphere. The local closure was used in a multi-species 10-moment fluid model to simulate Mercury's magnetosphere, specifically targeting the tightly coupled interior magnetosphere where collisionless magnetic reconnection is important in the magnetotail and magnetopause. \citet{dong_global_2019} utilised the local three-dimensional Landau closure of \citet{wang_comparison_2015} in order to include key kinetic effects while keeping computational load low, a key consideration for large-scale global simulations. The characteristic wavenumber required for the local approximation of the HP closure is defined as $k_s(\boldsymbol{x},t) = 10/d_s(\boldsymbol{x},t)$, where $d_s$ is the local inertial length of species $s$ as a function of position and time, allowing for the closure to capture the variation in inertial length across the entire Mercury system. \citet{dong_global_2019} deployed the closure in a multi-species fluid simulation and compared with \textit{in situ} data from the MESSENGER spacecraft, finding that the \citet{wang_comparison_2015} closure gave better agreement than a standard MHD approach.

In a parallel effort to extend the Hammett-Perkins closure to three dimensions, \citet{ng_simulations_2017} retained the full non-local HP form in a two-fluid, ten-moment model applied to magnetic reconnection. The closure, similarly to \citet{hammett_fluid_1990} is of the heat flux perturbation from equilibrium, extended to three dimensions:
\begin{align}
    q_{ijk}(\boldsymbol{x}) = n(\boldsymbol{x})\hat{q}_{ijk}(\boldsymbol{x}), \\
    \tilde{q}_{ijk} = -\text{i} \frac{v_t}{|k|} \chi k_{[i} \tilde{T}_{jk]},
\end{align}
where $\tilde{q}$ is the Fourier transform of $\hat{q}$ and $\tilde{T}$ is the Fourier transform of the deviation of the local temperature tensor from the mean. The coefficient $\chi$ was set as $\chi = \sqrt{4/9\pi}$ from a best fit of the diagonal $q_{iii}$ component. By keeping the fully non-local closure \citet{ng_simulations_2017} were able to include a description of phase mixing in the fluid model and test the ability of the model to accurately capture magnetic reconnection. \citet{ng_simulations_2017} performed simulations of magnetic island coalescence in order to test magnetic reconnection in a more complex problem than simpler cases, such as Hall sheet reconnection.

\citet{ng_improved_2020} subsequently took a similar route but used a local approximation of the HP closure extended to three dimensions. Unlike the \citet{wang_comparison_2015} local closure, whose $(P_{ij} - p\delta_{ij})$ form relaxes toward temperature isotropy, \citet{ng_improved_2020} took a local approximation to the heat flux directly, retaining anisotropy. This local approximation allows for a less computationally expensive closure without needing to approach pressure isotropy. \citet{ng_improved_2020} demonstrated the local closure in a simulation of the island coalescence problem to investigate magnetic reconnection, similarly to the approach using the non-local closure by \citet{ng_simulations_2017}, focusing on the ion diffusion region (the region over which ions become unbound to the magnetic field during the reconnection process) and the reconnection rate.

Many models have also been developed, similarly to the adiabatic fluid models, to include finite Larmor radius effects, which are not typically included by the earlier Landau fluid models. Including the finite Larmor radius effects requires the inclusion of nongyrotropic components of the second or third order moment tensors and can therefore increase the complexity of the equations considerably.

A Landau-fluid closure for the fourth-order moment in warm collisionless plasmas was constructed by \citet{goswami_landau_2005}, who incorporated FLR corrections by retaining the parallel and transverse heat-flux vectors and their coupling to nongyrotropic pressure at linear order. This linear treatment of nongyrotropic pressure and heat-flux terms, together with the assumption of small-amplitude perturbations on large spatial and temporal scales, holds only for weakly non-linear, large-scale dynamics. Within this regime, they showed that the model was able to reproduce the dispersion relations of all large-scale dispersive MHD waves in a warm collisionless plasma and to capture the mirror instability threshold with good accuracy. The closure strategy closely parallels that of \citet{snyder_landau_1997} for the “4+2” moment model, but they demonstrated that, for the $\tilde{r}_{\parallel \perp}$ component of the fourth-order moment, the two-pole Pad\'{e} approximation to the plasma response function employed by \citet{snyder_landau_1997} does not adequately represent the strong decay of its imaginary part, and they therefore adopted a three-pole Pad\'{e} approximation, improving the fidelity of the linear kinetic response at the cost of some additional analytical complexity.

A closely related FLR model, this time tailored to mirror-mode dynamics, was proposed by \citet{passot_fluid_2006}. Although valid only for low-frequency, large-scale dynamics under gyrotropic assumptions that neglect nongyrotropic pressures, the closure model successfully reproduced the mirror instability threshold and growth rates in agreement with linear kinetic theory. This work built directly on the linear-order nongyrotropic contributions introduced by \citet{goswami_landau_2005}, but focused more on mirror mode physics by emphasizing temperature-driven instabilities as opposed to the general dispersive waves considered by \citet{goswami_landau_2005}. The model offered a computationally efficient alternative to fully kinetic simulations for studying mirror mode generation and saturation in space plasma environments, despite limitations in resolving sub-proton-scale structures or strong turbulence.

A further advance came with \citet{sulem_landau_2015}, whose closure combined non-linear large-scale FLR effects with linear sub-ion-scale FLR and Landau-damping corrections from low-frequency linear kinetic theory. This hybrid approach produced a computationally tractable fluid description suitable for collisionless, low-frequency plasma turbulence, where it accurately captured the dispersion relations of kinetic Alfvén wave (KAW) turbulence and the mirror instability threshold. Although the model successfully combined physics-based non-linear estimates for large-scale nongyrotropic contributions with Pad\'{e}-approximated non-local Landau fluid operators, its reliance on low-frequency asymptotics and linear-kinetic foundations for damping terms confined it to weakly non-linear dynamics at scales above a fraction of the ion inertial length. Within this regime, however, the closure demonstrated clear improvements over prior linear-order models by retaining essential Hall-MHD and leading-order FLR physics while supporting temperature anisotropies through distinct parallel and perpendicular pressures.

A simplified version of the \citet{sulem_landau_2015} closure with a local Hilbert operator from \citet{passot_fluid_2014} and no FLR was used by \citet{finelli_bridging_2021} in a hybrid model of moderate-guide-field reconnection. The fluid closure was used in a hybrid model for the electron fluid and tested on simulations of moderate guide-field reconnection. They reported good overall agreement with a fully kinetic PIC reference, although the simplified (local Hilbert, no-FLR) closure does not reproduce all PIC features and behaves closer to an isothermal description in places.

Turning from Landau to cyclotron resonance, \citet{jikei_non-local_2021} defined a fluid model for collisionless magnetised plasmas that captures cyclotron-resonance effects. The closure was derived through a similar approach to that of \citet{hammett_fluid_1990} for the Landau resonance, applied instead to cyclotron resonance, which is of particular interest for space and astrophysical plasma where plasma beta can be large. The cyclotron resonance effects act on the transverse components of the higher-order moment tensors, as opposed to Landau resonance, which acts on the longitudinal components, relative to the electromagnetic perturbations. \citet{jikei_non-local_2021} therefore introduced a HP-like phase mixing closure of the transverse components of the heat flux tensor.

The same authors then compared the CRC model with a CGL-like adiabatic fluid model --- including FLR corrections and a Landau-fluid closure for the longitudinal mode --- in \citet{jikei_critical_2022}. The CGL-like model involves a closure of the heat flux, using the \citet{snyder_landau_1997} '3+1' moment closure of the parallel and perpendicular heat flux. The CGL-like model also includes FLR corrections via the inclusion of the nongyrotropic pressure components to linear order. The two fluid models were compared in simulations of the parallel firehose instability.

In summary, Landau fluid models, beginning with the work of \citet{hammett_fluid_1990}, incorporate the effects of phase mixing, allowing closure relations to capture collisionless damping within a fluid framework. This enables the fluid models to reproduce kinetic phenomena such as the mirror instability, wave-particle interactions, and aspects of magnetic reconnection more accurately than traditional fluid models. Although non-local formulations of the Landau fluid models can be computationally demanding, they remain significantly more efficient than fully kinetic treatments, and simplified local approximations offer a balance between physical fidelity and computational cost.

Despite their advantages, Landau fluid models also have notable limitations. Most closures are derived from the linear kinetic response function and therefore neglect non-linear effects that can be essential in strongly driven or turbulent regimes. Furthermore, the common assumption of a leading-order Maxwellian or bi-Maxwellian equilibrium restricts their applicability in scenarios where particle distributions deviate substantially from these forms, such as in reconnection exhausts or strongly anisotropic plasmas.

\subsubsection{Gyrofluid Models}

Gyrofluid models provide a reduced-fluid representation of kinetic plasma dynamics that is derived from the gyrokinetic equation and therefore sits within the broader plasma closure problem. In this framework, one starts from the Vlasov-Maxwell system and introduces a small parameter $\epsilon$, typically taken as the ratio of fluctuation frequency to gyrofrequency, the ratio of perpendicular fluctuation scale to the Larmor radius, or the ratio of $\boldsymbol{E} \times \boldsymbol{B}$ drift speed to the thermal speed, so that $\epsilon \sim \omega/\Omega \sim k_{\parallel}/k_{\perp} \sim v_{\boldsymbol{E}\times \boldsymbol{B}}/v_{th} \ll 1$. Under this ordering, the plasma is strongly magnetized with anisotropic fluctuations satisfying $\omega \ll \Omega$, $k_{\perp} \rho \sim 1$, and $k_{\parallel} \ll k_{\perp}$, and the fast gyromotion can be averaged over while retaining finite Larmor radius effects. The gyrokinetic equation is then obtained by averaging over the gyrophase of the particle motion about magnetic field lines and transforming to gyrocenter (or guiding-center) coordinates, which reduces the phase space dimensionality from six to five by removing the gyrophase angle while keeping the gyrocenter position, parallel velocity, and magnetic moment as independent variables. In physical terms, this procedure is equivalent to treating charged rings as quasiparticles following the motion of guiding centers, with the rapid cyclotron motion removed and included instead through the gyroaveraging and finite Larmor radius operators.

Gyrokinetic models constructed in this way are primarily intended to describe microturbulence driven by drift-wave-like and ion-temperature-gradient (ITG) instabilities in strongly magnetized plasmas, and they now serve as the standard theoretical framework for turbulent transport in fusion devices and, increasingly, in space and astrophysical plasmas where similar scale separations and strong magnetization have been observed. The gyrofluid approach then takes velocity moments of the gyrokinetic equation in gyrocenter coordinates, yielding a hierarchy of fluid-like equations for gyrocenter density, parallel flow, temperatures, and higher-order moments, coupled through gyroaveraged field equations that, in principle, retain finite Larmor radius corrections to all orders. Within this hierarchy, gyrofluid models face a closure problem analogous to that of conventional fluid descriptions: evolution equations are available only for a finite set of low-order moments, while higher-order moments—and the associated phase mixing and non-local kinetic responses—must be modelled. HP-like Landau-fluid closures often combined with finite Larmor radius operators inherited from the gyrokinetic derivation, are widely used in gyrofluid models to reproduce linear Landau damping and collisionless phase mixing with good fidelity while keeping the equations tractable for non-linear turbulence simulations. As a result, gyrofluid models can incorporate many of the closure strategies discussed earlier, such as Landau-fluid and FLR-Landau-fluid closures, within a framework that is consistent with the gyrokinetic ordering and dimensionality reduction, making them particularly suitable for efficient modelling of microturbulent transport in fusion, space, and astrophysical plasmas.

The foundational non-linear gyrofluid formulation is due to \citet{brizard_nonlinear_1992}, who began from the gyrokinetic Vlasov equation for a nonuniform, strongly magnetised plasma with low-frequency fluctuations and electromagnetic perturbations. They took gyrocenter fluid moments (density, parallel velocity, pressures, etc. of gyrocenters) of the gyrokinetic equation to obtain a hierarchy of gyrofluid equations that automatically include FLR corrections and gyroviscous effects. These non-linear gyrofluid equations couple gyrocenter moments to the field equations (Poisson/quasi-neutrality and Amp\`{e}re-like relations) with finite-$\rho_i$ effects retained in the non-linear terms, so the model can describe microturbulence and transport driven by drift/Alfv\'{e}nic instabilities in a magnetized plasma. They showed that in appropriate limits one recovers FLR-corrected reduced fluid equations, including reduced Braginskii equations and ideal reduced MHD when FLR is neglected.

A systematic gyrofluid model derived directly from the electrostatic gyrokinetic equation was given by \citet{dorland_gyrofluid_1993}, incorporating Landau-fluid closures to capture collisionless damping. Their work extends the Landau-fluid closure approach of \citet{hammett_fluid_1990, hammett_fluid_1992} to the gyrokinetic framework, allowing a consistent treatment of microturbulence and linear phase mixing within a fluid model. These equations naturally include finite Larmor radius (FLR) effects through the gyrokinetic derivation while employing HP closures to represent phase mixing and Landau damping, reproducing kinetic damping rates with high fidelity. A key contribution of this study is the formulation of a generalizable, systematic procedure for constructing fluid models from the gyrokinetic hierarchy at any chosen order of velocity moment. This enables the derivation of increasingly accurate models by retaining additional higher-order moments as needed for specific physical regimes. Importantly, \citet{dorland_gyrofluid_1993} addressed two distinct closure problems that arise in this approach: one associated with the typical fluid moments of the distribution function and another with the moments of the gyroaveraged operator. The former is treated using Landau-fluid closures that separately account for parallel and perpendicular dynamics, while the latter employs a guiding-center-based closure scheme.

\citet{passot_gyrofluid_2018} later reduced the \citet{brizard_nonlinear_1992} gyrofluid model asymptotically for low electron plasma beta, strong guide field, and anisotropic turbulence $k_{\perp} \ll k_{\parallel}$, keeping ion FLR, electron inertia and $B_{\parallel}$ fluctuations. The reduction yields a two-field model (a flux function and a density/parallel magnetic fluctuation variable) that describes shear Alfv\'{e}n waves at MHD scales and dispersive kinetic Alfv\'{e}n waves down to below the electron inertial length $d_{e}$. It was demonstrated that this reduced system is a noncanonical Hamiltonian, with a well-defined free energy, Casimirs and generalized cross-helicity, which then constrain non-linear dynamics and turbulence cascades. This two-field model has been applied to the numerical study of solar-wind kinetic Alfv\'{e}n-wave turbulence, including inverse cascade and magnetic-vortex formation \citep{miloshevichInverseCascadeMagnetic2021}.

\citet{zhou_electron_2023} built a gyrofluid model on the Kinetic Reduced Electron Heating Model (KREHM) of \citet{zocco_reduced_2011}. The KREHM framework is a rigorous, asymptotically reduced model derived from the gyrokinetic equation under the assumption of low electron plasma beta. In their work, they obtained a set of fluid equations from the KREHM formulation and achieved closure by expressing the reduced electron distribution function, $g_e$, as an expansion in Hermite polynomials. Here, $g_e$ represents the part of the electron distribution function that encapsulates information about the second- and higher-order velocity-space moments of $\delta f_e$, the perturbed electron distribution function. The use of a Hermite expansion for $g_e$ follows the approach established by \citet{zocco_reduced_2011}. To investigate the resulting model dynamics, they performed numerical simulations of the KREHM system using the Viriato code \citep{loureiro_viriato_2016}. 

The approach of expanding the electron distribution function in Hermite polynomials, which underpins both the KREHM formulation and the Viriato code, exemplifies a broader methodology championed by Loureiro and collaborators for bridging kinetic and fluid descriptions of magnetised plasmas. In a Hermite representation, the velocity-space dependence of the perturbed distribution function is decomposed onto a discrete set of orthogonal polynomials, so that each Hermite coefficient corresponds to a progressively higher-order velocity-space moment. Using the KREHM framework, \citet{zhou_electron_2023} showed analytically and numerically that efficient electron heating in sub-$\rho_i$ turbulence is primarily due to Landau damping of kinetic Alfv\'{e}n waves rather than Ohmic dissipation, and that this heating is concentrated around intermittent current sheets where advective non-linearities are locally suppressed and phase mixing can proceed unimpeded. In the same framework, \citet{zhou_spectrum_2023} investigated the turbulent cascade of kinetic Alfv\'{e}n wave turbulence using an isothermal fluid description. These results demonstrate the utility of the Hermite-gyrofluid methodology, as both a computational tool and an analytical framework, for studying the coupled position- and velocity-space dynamics of kinetic turbulence in weakly collisional plasmas.

Taken together, gyrofluid models have been successful within their domain of applicability. By combining the dimensionality reduction of the gyrokinetic ordering with Landau-fluid closures, they provide a set of equations that can reproduce linear kinetic damping rates with high fidelity \citep{dorland_gyrofluid_1993, hammett_developments_1993}, capture finite Larmor radius effects to good accuracy across a broad range of $k_{\perp} \rho$ \citep{dorland_gyrofluid_1993, beer_toroidal_1996}, and retain the essential non-linear couplings that drive microturbulence and transport \citep{beer_toroidal_1996}. Extensions to toroidal geometry have enabled direct simulations of ion-temperature-gradient-driven turbulence in realistic tokamak configurations \citep{beer_toroidal_1996}, and Hamiltonian gyrofluid formulations have furnished powerful analytical tools for studying energy conservation, non-linear cascades, and the structure of kinetic Alfv\'{e}n wave turbulence in space and astrophysical plasmas \citep{passot_gyrofluid_2018, passot_imbalanced_2019}.

However, gyrofluid models, like the other fluid models, are subject to several limitations. Gyrofluids are inherently constrained by the gyrokinetic ordering on which they are built, the assumptions $\omega \ll \Omega$, $k_{\perp} \rho \sim 1$, and $k_{\parallel} \ll k_{\perp}$ restrict the models to low-frequency, strongly magnetised, anisotropic fluctuations, and phenomena that violate this ordering lie outside their scope. Secondly, the transformation to gyrocenter (guiding-center) coordinates, while essential for averaging out the rapid gyromotion, introduces considerable algebraic complexity into the non-linear equations, particularly when electromagnetic perturbations, toroidal geometry, or multiple species are retained \citep{brizard_nonlinear_1992, beer_toroidal_1996}. Third, as \citet{dorland_gyrofluid_1993} emphasised, gyrofluid models generically face two distinct closure problems. A closure for the conventional fluid moment hierarchy, where the highest retained velocity-space moment must be approximated in terms of lower ones, and another for the moments of the gyroaveraging operator, which require a separate closure scheme typically formulated in guiding-center coordinates. Each of these closures introduces its own approximation errors. Fourth, the fluid closures most commonly employed in gyrofluid models are HP-type non-local Landau-fluid closures, and these therefore inherit the well-known difficulties of such closures: they are constructed from the linear kinetic response and so will not be fully valid in systems where non-linear effects modify the velocity-space dynamics relative to linear expectations \citep{schekochihin_phase_2016, shukla_learned_2022}. Although gyrofluid models have been shown to reproduce macroscopic transport quantities (such as the radial heat flux) with reasonable accuracy even when the underlying phase mixing rates are not perfectly captured \citep{shukla_learned_2022}, the reliance on linearly derived closures remains a fundamental limitation for non-linear applications.

\subsection{From Analytic Closures to ML Closures}

The preceding sections have traced the development of analytic closure models for collisionless and weakly collisional plasmas, from the earliest double-adiabatic (CGL) closure \citep{chew_boltzmann_1956} through successive generations of finite Larmor radius corrections, Landau-fluid models, and gyrofluid formulations. In this development, each new closure has extended the fluid description by incorporating additional kinetic effects (pressure anisotropy, collisionless damping, finite-gyroradius averaging) into moment hierarchies of increasing sophistication. At the same time, the variety of available closures causes a fundamental difficulty at the heart of the closure problem: there is no unique or universally optimal way to truncate the moment hierarchy, and the choice of closure is always a compromise between physical fidelity and practical tractability.

On one hand, the most physically faithful closures tend to be the most computationally demanding. Non-local Landau-fluid closures require the evaluation of Hilbert transforms in wavenumber space, which can be expensive in configuration-space codes and are difficult to generalise to inhomogeneous geometries \citep{hammett_fluid_1990, hammett_fluid_1992}. Higher-order moment models that evolve six, eight, or ten fluid quantities per species improve accuracy but multiply the number of coupled equations of the resulting system \citep{wang_comparison_2015, ng_improved_2020}. On the other hand, simpler local closures and low-moment truncations sacrifice important kinetic physics. The CGL closure neglects heat flux entirely, and local approximations to Landau damping distort the dispersion and damping of kinetic-scale waves. The results is a wide variety of analytic closures, each optimised for a particular regime, geometry, or class of instability, with no single model spanning the full range of conditions encountered in space, astrophysical, and laboratory plasmas.

Beyond these practical trade-offs, the analytic closures reviewed above share a set of more fundamental limitations that constrain their applicability regardless of the number of moments retained. Finite Larmor radius models are built on an asymptotic expansion in the ratio of the Larmor radius to the characteristic gradient scale, $k_{\perp} \rho \ll 1$ or $k_{\perp} \rho \sim 1$ depending on the ordering, and they thus lose accuracy when fluctuations reach scales comparable to or smaller than the Larmor radius. This is one reason why standard MHD, which omits FLR corrections altogether, can serve as an approximation for the large-scale dynamics of collisionless plasmas. On scales much larger than the ion gyroradius, FLR effects are small corrections to the leading-order force balance, and the MHD equations capture the dominant physics. Conversely, the same expansion makes FLR-based closures difficult to apply in weak-guide-field configurations, where the Larmor radius can become comparable to or exceed the relevant fluctuation scales, invalidating the underlying ordering.

Additionally, the Landau-fluid closures are typically derived from the linear kinetic response function by matching a Pad\'{e} approximant to the linear plasma dispersion function in the low- or high-frequency asymptotic limit \citep{hammett_fluid_1990, hammett_fluid_1992}. This results in closures that are constructed to reproduce the correct damping rates and dispersion relations for small-amplitude perturbations about a Maxwellian (or bi-Maxwellian) equilibrium, but there is no guarantee that they remain accurate in fully non-linear regimes where the distribution function departs substantially from equilibrium. As discussed in the context of gyrofluid models, turbulent systems can exhibit phase mixing rates that are heavily suppressed relative to the linear predictions, owing to non-linear effects such as stochastic plasma echoes and the coupling between phase mixing and anti-phase mixing perturbations \citep{schekochihin_phase_2016, shukla_learned_2022}. Third, the assumption of a leading-order Maxwellian or bi-Maxwellian distribution restricts the validity of most closures to plasmas that remain close to local thermodynamic equilibrium. In environments such as reconnection exhausts, shock transition layers, or strongly driven turbulence, where the distribution function can develop beams, flat-top profiles, or other strongly non-Maxwellian features, the underlying expansion on which the closure is based may no longer be justified.

These limitations persist throughout all analytical closure development, and will continue to do so without a new fundamental development in theory. It is in addressing these limitations that machine learning methods may offer an alternative approach. The term `machine learning' here encompasses a broad family of data-driven methods, from neural network regression to equation discovery techniques such as sparse regression, symbolic regression, and genetic programming (discussed in detail in Section~2.2.2).

Neural networks are capable of representing arbitrarily complex non-linear mappings between input and output fields, at a fraction of the computational cost of fully-kinetic simulations. This combination of expressivity and computational efficiency makes them natural candidates for the closure problem. Rather than deriving an analytic relation between the highest retained moment and the lower-order quantities, one can instead train a neural network to learn this relation directly from high-fidelity kinetic simulation data. Because the network is not constrained by the assumptions of linear response, Maxwellian equilibrium, or a particular asymptotic ordering, it can in principle capture closure relations that hold in regimes where analytic models fail, including the fully non-linear, non-equilibrium conditions that are most relevant to many space and astrophysical plasmas.

A complementary and in some respects more interpretable class of data-driven methods is provided by equation discovery techniques, which aim to recover explicit symbolic expressions for the closure relation. The paradigmatic example is the sparse identification of non-linear dynamics (SINDy) framework \citep{brunton_discovering_2016}, which constructs a library of candidate non-linear terms and uses sparsity-promoting regression to select the minimal subset that best fits the training data \citep{rudy_data-driven_2017}. The result is a parsimonious partial differential equation whose terms are directly interpretable in terms of the underlying physics, an advantage over opaque neural network models when the goal is not merely predictive accuracy but physical insight.

Early work in the neural network direction has already demonstrated the feasibility of the approach. \citet{ma_machine_2020} showed that multilayer perceptrons, convolutional neural networks, and Fourier-space networks can all learn the non-local Hammett-Perkins closure in configuration space, reproducing its characteristic wavenumber-dependent structure with high accuracy. Subsequent studies have extended this to more complex closures and integrated the resulting surrogate models into fluid simulations, demonstrating that neural network closures can be coupled to fluid solvers and run stably in time \citep{wang_deep_2020, maulik_neural_2020}. More recently, \citet{huang_machine-learning_2025} used a Fourier neural operator to learn a heat flux closure from first-principles Vlasov simulation data and showed that the resulting fluid model could reproduce the full non-linear evolution of Landau damping, a result that no analytic fluid closure had previously achieved. On the equation discovery side, \citet{alves_data-driven_2022} demonstrated that an integral formulation of sparse regression can recover the full hierarchy of plasma fluid models directly from particle-in-cell simulation data. \citet{donaghy_search_2023} subsequently applied SINDy to seek an improved symbolic closure for a ten-moment multi-fluid model, using fully kinetic simulation data from a Harris-sheet reconnection scenario. These developments, spanning both neural network and equation discovery paradigms, and the broader programme of data-driven closure modelling that they represent, form the subject of the following sections.

{\small
\begin{longtable}{p{2.4cm} p{1.7cm} p{3.0cm} p{5.3cm}}
\caption{Summary of analytical plasma closure models discussed in this review, organised by closure class. ``Closure Class'' indicates the type of fluid model and the order at which the fluid hierarchy is truncated: 5-moment (scalar pressure), 10-moment (full pressure tensor), or higher-order closures such as the heat flux or the 4th order moment.}
\label{tab:closures} \\

\toprule
\thead{Reference} & \thead{Closure\\Class} & \thead{Target Regime/\\Phenomenon} & \thead{Key Points/\\Limitations} \\
\midrule
\endfirsthead

\multicolumn{4}{c}{\small\itshape Table~\ref{tab:closures} continued from previous page} \\
\toprule
\thead{Reference} & \thead{Closure\\Class} & \thead{Target Regime/\\Phenomenon} & \thead{Key Points/\\Limitations} \\
\midrule
\endhead

\midrule
\multicolumn{4}{r}{\small\itshape Continued on next page} \\
\endfoot

\bottomrule
\endlastfoot

% ===== ADIABATIC MODELS =====
\multicolumn{4}{l}{\textit{Adiabatic Fluid Models}} \\
\midrule

\citet{chew_boltzmann_1956}
  & Adiabatic, 5-moment
  & General collisionless; ions only
  & Introduces $p_\parallel$, $p_\perp$ anisotropy via conservation of adiabatic invariants; $q=0$ closure; no electron dynamics, no FLR \\
\addlinespace

\citet{abraham-shrauner_propagation_1967, abrahamshrauner_doubleadiabatic_1968, abraham-shrauner_small_1973}
  & Adiabatic (polytropic), 5-moment
  & Plasma wave propagation
  & Generalised polytropic indices for anisotropic pressure; bridges MHD and CGL limits; no kinetic damping \\
\addlinespace

\citet{ballai_linear_2002}
  & Adiabatic (polytropic), 5-moment
  & Linear and non-linear waves in a slab
  & Two-index polytropic law; derives wave dispersion relations; limited to slab geometry \\
\addlinespace

\citet{le_equations_2009}
  & Adiabatic (electron trapping), 5-moment
  & Guide-field magnetic reconnection
  & Derives $p_\parallel(n,B)$, $p_\perp(n,B)$ from electron trapping analysis; recovers CGL invariants in zero heat-flux limit; limited to guide-field reconnection geometry \\
\addlinespace

\citet{yajima_effect_1966}
  & Adiabatic + FLR, 5-moment
  & Small-amplitude oscillations
  & First-order FLR correction to CGL; derives modified dispersion relations; limited to linearised dynamics \\
\addlinespace

\citet{hazeltine_local_2013}
  & Adiabatic + FLR, 5-moment
  & General magnetized plasma (with collisions)
  & Thermodynamic derivation of internal energy with CGL, gyroviscous, and FLR contributions; includes collisional effects \\

\addlinespace
\midrule
% ===== LANDAU FLUID MODELS =====
\multicolumn{4}{l}{\textit{Landau Fluid Models}} \\
\midrule

\citet{hammett_fluid_1990, hammett_fluid_1992}
  & Landau fluid (non-local), heat flux
  & Electrostatic Landau damping
  & Pad\'e approximation of plasma response function; $1/|k|$ non-locality; captures linear Landau damping; limited to linear, electrostatic regime \\
\addlinespace

\citet{snyder_landau_1997}
  & Landau fluid (non-local), ('3+1'), ('4+2')
  & Collisionless bi-Maxwellian; mirror instability
  & Electromagnetic, drift-kinetic hierarchy with HP-type closures; correctly reproduces mirror instability threshold and MHD-wave dispersion; includes collisional limit \\
\addlinespace

\citet{wang_comparison_2015}
  & Landau fluid (local HP approx.), 5- and 10-moment
  & Collisionless reconnection (Harris sheet)
  & Local approximation of HP closure via relaxation to temperature isotropy and characteristic wavenumber $k_{0}$; 10-moment model produces results comparable to PIC for reconnection; computationally efficient \\
\addlinespace

\citet{ng_simulations_2017}
  & Landau fluid (non-local HP), 10-moment
  & Reconnection; anisotropic heat flux
  & Extends non-local HP closure to 3D for both ions and electrons; tested on island coalescence problem \\
\addlinespace

\citet{ng_improved_2020}
  & Landau fluid (local HP approx., 3D), 10-moment
  & Island coalescence; reconnection
  & Replaces $1/|k|$ with single $k_0$ for local closure; without relaxation to temperature isotropy; improved performance over other local closures; suited for large-scale simulations \\
\addlinespace

\citet{dong_global_2019}
  & Landau fluid (local HP), 10-moment
  & Mercury's magnetosphere (global)
  & Full global magnetosphere simulation using Wang \textit{et al.}\ (2015) closure; improves upon Hall-MHD with off-diagonal pressure terms \\
\addlinespace

\citet{goswami_landau_2005}
  & Landau fluid (non-local)+ FLR, 4-th order moment
  & MHD wave dispersion; general low-frequency
  & 3D Landau 2-fluid closure assuming bi-Maxwellianity; includes FLR corrections up to 2nd order and nongyrotropic pressure/heat-flux at linear level; reproduces all MHD wave dispersion relations \\
\addlinespace

\citet{sulem_landau_2015}
  & Landau fluid (non-local) + FLR, extended moment
  & Kinetic Alfv\'en wave turbulence; mirror modes
  & Merges non-linear FLR dynamics with linear sub-ion Landau kinetics; reproduces kinetic Alfv\'en wave dispersion and mirror instability; limited by low-frequency, linear-kinetic assumptions and non-local Landau operator approximations \\
\addlinespace

\citet{finelli_bridging_2021}
  & Landau fluid (simplified, local), heat-flux
  & Moderate guide-field reconnection
  & Simplified Sulem-Passot closure (local Hilbert operator approximation, no FLR); hybrid model for electron fluid; recreates reconnection dynamics at reduced computational cost \\
\addlinespace

\citet{jikei_non-local_2021}
  & Landau fluid + FLR, heat-flux
  & Electromagnetic ion cyclotron wave resonance
  & FLR Landau closure capturing cyclotron resonance; comparison of CGL+FLR+Landau fluid vs.\ alternative model \\

\addlinespace
\midrule
% ===== GYROFLUID MODELS =====
\multicolumn{4}{l}{\textit{Gyrofluid Models}} \\
\midrule

\citet{brizard_nonlinear_1992}
  & Gyrofluid, non-linear (gyrokinetic moments)
  & Drift/Alfv\'enic microturbulence; strongly magnetised plasma
  & Gyrocenter fluid moments of the non-linear gyrokinetic Vlasov equation; automatically retains FLR and gyroviscous effects; foundational non-linear gyrofluid formulation. \\
\addlinespace

\citet{dorland_gyrofluid_1993}
  & Gyrofluid, heat flux (Landau-fluid)
  & Electrostatic ion-temperature gradient instability/drift-wave microturbulence
  & Systematic gyrofluid hierarchy derived from the electrostatic gyrokinetic equation with Landau-fluid closures; addresses two distinct closure problems (fluid moments and moments of the gyroaveraging operator); separate parallel/perpendicular closures. \\
\addlinespace

\citet{passot_gyrofluid_2018}
  & Gyrofluid, extended moment
  & Low-frequency collisionless dynamics
  & Contemporary gyrofluid closure of gyrokinetic hierarchy; includes FLR and Landau damping; applicable to sub-ion-scale turbulence \\
\addlinespace

\citet{zhou_electron_2023}
  & Gyrofluid (+ Hermite), extended moment
  & Electron heating via kinetic Alfv\'en wave Landau damping
  & Reduced strong-guide-field gyrofluid model with ion FLR + electron inertia; optionally couples to drift-kinetic electrons expanded in Hermite basis; demonstrates energy transfer to electrons via Landau damping of kinetic Alfv\'en waves. \\

\end{longtable}
}

\section{Machine Learning}

The purpose of applying ML techniques to the plasma closure problem is twofold: Including higher-order effects into large-scale simulations and improving the computational cost and speed of simulations. Similar work has been ongoing in computational fluid mechanics for some time, where the aim is primarily to emulate fluid models using ML to obtain fast-running models that can be employed in near-real time \citep{vinuesa_enhancing_2022}.

In the interest of readers who may not be familiar with Machine Learning terminology and methodology, set out here are some of the key concepts needed for the application of ML techniques to the plasma closure problem. Similarly to the previous section, this will be kept relatively brief. There are many good textbooks and reviews of machine learning methodology in general, and specially for the application to physics, \citet{mehta_high-bias_2019} gives a (relatively) compact review of machine learning techniques of relevance to physics. \citet{Goodfellow-et-al-2016} gives a very in depth and comprehensive textbook covering deep learning, the subfield of machine learning dedicated to multi-layer neural networks. We also include a glossary of ML terminology in Appendix A that is used throughout the subsequent discussion for quick reference.

Broadly, ML problems can be divided into three main categories relevant here (reinforcement learning, a fourth paradigm, is not used in the closure studies reviewed).
\begin{itemize}
    \item Supervised learning: The ML model is trained on labelled data where each (multidimensional) input data point is paired with an output data point (‘label’ or `ground truth'). Supervised learning tasks typically consist of regression (continuous labels) and classification (discrete labels).
    \item Unsupervised learning: The ML model is trained on unlabelled data, where the goal is to extract a pattern or underlying behaviour from the data without any information about a specific solution. Examples of unsupervised learning tasks typically include clustering and dimensionality reduction.
    \item Semi-Supervised learning: The ML model is trained with a mixture of labelled and unlabelled data. An example of a semi-supervised learning technique is the physics-informed neural network training method, where labelled data can be combined with a physics-informed loss function. Other examples include tasks where labelled data is sparse or time-consuming to generate and methods are used to train on both the labelled and unlabelled data, such as in computer vision tasks for classifying images, where unlabelled data can be utilised for segmenting images and labelled data for classifying of segments.
\end{itemize}

The plasma closure problem has so far been approached as a supervised learning problem, where labelled training data is generated through high-resolution kinetic simulations directly, or from a known analytic closure equation. A semi-supervised approach is also applicable, however, utilising techniques such as physics-informed neural networks \citep{raissi_physics-informed_2019}.

The goal, then, is to choose the most appropriate techniques from the field of machine learning to train a model for the plasma moment closure able to capture kinetic effects and to be integrated into large-scale plasma simulations without incurring a steep computational cost. Although the plasma community has only recently started to appreciate the potential of using machine learning in the closure problem, in this review, we aim to collate several preliminary works. We hope the review will inform future efforts and highlight possible research directions.

\subsection{Why ML?}

A pertinent question to ask is why a data-driven approach to improving plasma closure models is appropriate and useful. Some of the rationale is related to the relative success of employing machine learning in the neighbouring field of neutral fluid dynamics (including turbulence)  \citep{li_graph_2022, cai_physics-informed_2021, brunton_machine_2020}.

Inherently, by truncating the system of equations, we are making an approximation to describe a reduced model of the plasma. In performing this approximation, the model loses detail, rendering simple plasma fluid models incapable of capturing kinetic effects. An approximation of this kind is inevitably a trade-off between accuracy and computational cost. By utilising a neural network, we can take advantage of its 'universal approximator' property to learn an approximate closure that could remain close to the fully kinetic model, while reducing the computational cost of modelling \citep{hornik_multilayer_1989}. In other words, if we accept the fundamental assumption of any fluid model that a closure exists and that it preserves, to some level of accuracy, the dynamics of low-order moments, then a neural network should be able to approximate such a closure.

A significant benefit of the neural network approach is in the low computational cost of making predictions with a trained neural network (and the accelerated inference time from highly parallelised frameworks operating on GPUs) relative to the complexity that the neural network can achieve. This advantage in computational cost is most pronounced relative to fully kinetic simulations. However, some approaches using dense, spatially non-local neural networks may entail a numerical complexity that could be impractically high within a fluid solver. Analytical expressions obtained via sparse regression are likely faster to evaluate than multi-layered neural networks. However, neural networks will typically be able to model arbitrary non-linearities, whereas the analytic expressions from sparse regression will be bounded by the candidate terms in the library, which are chosen \textit{a priori}.

In contrast to the 'black-box' approach of neural networks, multiple attempts are being made at using equation discovery methods, such as sparse and symbolic regressions, to find governing equations of systems. By utilising these equation discovery methods, an analytic form of the closure relation could be found, purely from a data-driven approach. Such analytic closures could aid in the development of theoretical work, as well as provide more stable models for forecasting and simulation.

\subsection{ML Models}

\subsubsection{Neural Networks}

Many recent studies aimed at improving plasma closure are utilising neural networks. The subject of neural networks and their applications (or deep learning) is large, and due to the current popularity of the field, it is expanding rapidly. For this review, an outline of the relevant neural network architectures will be given to provide extra context for some of the techniques discussed.

A neural network is a machine learning model that can approximate any continuous function to any desired level of accuracy, according to the universal approximation theorem \citep{hornik_multilayer_1989}. Neural networks, initially inspired by the human brain, are made up of 'neurons', arranged in layers, with neurons in subsequent layers connected via tunable parameters known as 'weights'. The individual neurons in neural networks can be represented as a simple linear combination of connected inputs and a non-linear 'activation function'.

\begin{equation}
    z \: = \sigma ( \mathsfbi{W}\boldsymbol{x} + \boldsymbol{b} )
\end{equation}
where $z$ is the output of the neuron, $\boldsymbol{x}$ is the input vector of the neuron, $\mathsfbi{W}$ is the matrix of weights representing the neural network layer, $\boldsymbol{b}$ is the bias vector, and $\sigma$ is the non-linear activation function. Typically, only $\mathsfbi{W}$ and $\boldsymbol{b}$ are trainable parameters of the neural network. The optimal weights and biases are estimated by solving an optimization problem that minimizes a loss function, defined as the error between the neural network output and the corresponding ground truth. Neural networks are often trained using the backpropagation algorithm, which enables a version of gradient descent to be applied efficiently to each trainable parameter of the network.

An advantage of the neural network approach is the ability for the model to be trained on fully kinetic simulation data, without any prior knowledge of the governing models. Neural networks are 'universal approximators', meaning that theoretically, a neural network with enough layers and neurons, and with enough available training data, can approximate the input data to an arbitrary level of accuracy \citep{hornik_multilayer_1989}. This characteristic of neural networks allows them to be applied to situations where little is known about the underlying model, which could be the case for some high-dimensional kinetic simulations, where an analytic closure model that can capture the dynamics is not available in closed form.

Physics-informed neural networks (PINNs), introduced by \citet{raissi_physics-informed_2019}, provide a framework for embedding the governing physics of a system directly into the architecture and training process of neural networks. Unlike conventional data-driven networks, PINNs enforce physical consistency by incorporating the governing equations, such as partial differential equations (PDEs), conservation laws, or constitutive relations—into the loss function.

This is achieved by evaluating the residuals of the governing equations at a set of collocation points distributed throughout the training domain. At each collocation point, the neural network predictions are inserted into the differential equations, and the residuals (i.e., deviations from the physically exact relations) are computed. These residuals form the physics-based loss term, which penalizes solutions that violate the governing equations during training. Hence, the total loss typically combines data mismatch terms and physics constraints, guiding the model toward both empirical accuracy and physical validity. The collocation points can differ from labelled input data, allowing deviation from the physics to be penalized at points in the domain without labelled data, making the PINNs training framework semi-supervised (or fully-unsupervised if only using a physics loss-term and no data-loss terms).

A key advantage of PINNs lies in their use of automatic differentiation (AD) to compute spatial and temporal derivatives of the network output with respect to its inputs. AD enables precise evaluation of differential operators without discretization error. PINNs represent a change in training methodology, rather than a different architecture of neural network, and thus a PINN training loss can be applied to many different possible network architectures.

The form of the physics-informed loss depends on the underlying system and the choice of physical constraints. For instance, when the PDE of the system is directly embedded as part of the loss, it acts as a regularization term that biases the network toward solutions lying within the manifold of admissible physical states \citep{karniadakis_physics-informed_2021}. Extensions of this approach have been developed to address challenges such as stiff and ill-conditioned systems and multi-scale dynamics \citep{krishnapriyan_characterizing_2021, wang_understanding_2021} and noisy or sparse data \citep{yangBPINNsBayesianPhysicsinformed2021}, further broadening the applicability of PINNs across scientific and engineering domains.

A more recent variation of the PINN has been developed by \citet{yu_gradient-enhanced_2022}, the gradient-enhanced PINN (gPINN), which includes the gradients of the physics loss terms as additional constraints during the training of the network.

Neural operators, developed by \citet{kovachki_neural_2024}, are networks that, instead of approximating a function, seek to approximate an operator representing the mapping between the input and output function spaces. Let $\mathcal{A} \in \mathbb{R}^{d}$ and $\mathcal{U} \in \mathbb{R}^{d'}$ be two Banach spaces of continuous functions. $\mathcal{G} : \mathcal{A} \rightarrow \mathcal{U}$ is the operator mapping between these two spaces. A neural operator attempts to approximate this operator mapping via the iterative update of $\mathcal{G}'$ \citep{kovachki_neural_2024}:
\begin{equation}
    \mathcal{G}' = \mathcal{Q} \circ \sigma_{T} (\mathsfbi{W}_{T-1} + \mathcal{K}_{T-1} + \boldsymbol{b}_{T-1}) \circ \cdots \circ \sigma_{1}(\mathsfbi{W}_1 + \mathcal{K}_1 + \boldsymbol{b}_1) \circ \mathcal{P},
    \label{eq:neuralop}
\end{equation}
where $\mathcal{Q}$ and $\mathcal{P}$ are the local projection and lifting mappings respectively. These are required as inputs will be finite-dimensional vectors of 'sensor' points of the continuous input functions, and thus input and output dimensions can differ. $\sigma_t$ are activation functions, $\mathsfbi{W}_t$ are local linear operators, $\boldsymbol{b}_t$ are bias functions and $\mathcal{K}_t$ are integral kernel operators. $\mathsfbi{W}_t$, $\boldsymbol{b}_t$ and $\mathcal{K}_t$ are learnable. Interestingly, for the case of learning the solution operator of a linear PDE, the integral kernel operator can be written as the Green's function of the PDE, however this is not always the case for more general operator mappings. From this explicit neural operator it is clear how the structure is analogous to the typical neural network architecture, extended to an operator framework.

Many variations of the neural operator have been developed, such as the Deep Operator Network (DeepONet) from \citet{lu_deeponet_2020} and the Fourier Neural Operator (FNO) from \citet{li_fourier_2021} A variation of the neural operator, including the principles of PINNs, has also been developed by \citet{li_physics-informed_2023}, the physics-informed neural operator (PINO). In principle, neural operators improve upon neural networks by being better adapted to including multiple initial conditions in their training, whereas a physics-informed neural network would need to be re-trained for each set of different initial and boundary conditions when approximating a system. Neural operators have also been shown to have a 'discretisation invariance' allowing training to be conducted on solutions obtained on a coarser discretisation to be later generalized on finer resolutions. Proof of the neural operator discretisation invariance is set out in Appendix E of \citet{kovachki_neural_2024}.

Another benefit, particularly applicable to neural operators and physics-informed neural networks, is the potential for greater adaptability and generalisation, provided the model is trained on sufficiently varied data. While well-generalisable models can be difficult to train, once trained, a neural network model could theoretically generalise well to different conditions of the system, a contrast to the analytic closure approach, which typically requires a different closure model for each regime and phenomena being investigated. The accuracy and generalisability of neural networks does however depend on the training data, and while a neural network trained on a variety of different conditions may be able to generalise within the subset of parameter space covered by the training data, generalisability beyond the training data will not be possible. Analytic closures obtained through sparse regression could potentially have better ability to generalise and to capture extremes that the training data used for neural network closures may not sufficiently represent.

The adaptability of these methods to multiple different plasma phenomena, however, has not been fully investigated as of yet, with only a few studies, such as the work by \citet{wei_data-driven_2023} and \citet{miloshevich_electron_2025}, attempting to investigate the ability for neural network closures to generalise to multiple different simulation datasets.

Neural networks, while promising, do suffer from some significant disadvantages. A primary concern for physics is the lack of interpretability of the neural network models. Extracting meaning from the models is increasingly difficult if the entire system of fluid equations is being approximated by a model, as is the case for some studies. Neural networks can also be subject to problems of accumulating errors and numerical instability when results are propagated in time \citep{frezat_posteriori_2022}, either through numerical schemes when the closure is included in a plasma model, or if using a time-dependent neural network, such as a recurrent neural network or a transformer. This is due to the small errors that will be present in the neural network approximation after training, which will eventually grow over time into large errors and possibly cause the growth of instabilities in the simulation.

The rapid progress in neural network research has also resulted in a large set of possible architectures and, therefore, choices that a researcher must make when finding the best solution to their problem. It is often not obvious which type of neural network will work best for the plasma closure problem, and the secondary issue of choosing the optimal hyperparameters for the neural network could also present some issues for future work in modelling the plasma closure relation. Much of the research in applying neural networks to the plasma closure has been concerned with finding which neural network architecture will be most suitable for the given regime and conditions. The work from \citet{ma_machine_2020} and \citet{maulik_neural_2020} represents early investigations into how the different neural network types perform for the given closure model.

\subsubsection{Equation Discovery}

An alternative to learning closure relations as numerical surrogates is to discover them as explicit mathematical expressions directly from data. The field of equation discovery, broadly reviewed by \citet{camps-valls_discovering_2023}, encompasses a family of techniques that seek to recover the governing equations of a dynamical system from observations, without assuming a specific functional form a priori. In the context of the plasma closure problem, equation discovery methods aim to identify closed-form relations for the higher-order moments expressed as functions of the resolved fluid variables and their gradients. Because the output is an interpretable analytical expression rather than a trained network, the resulting closure can be inspected, simplified, and embedded directly into existing fluid solvers with minimal additional computational cost. Two broad strategies are in common use: sparse regression, which selects terms from a predefined library, and symbolic regression, which searches over a wider space of mathematical expressions, typically by means of genetic algorithms. We discuss each in turn.

Sparse regression methods can infer the governing partial differential equations (PDEs) of dynamical systems by identifying the coefficients of a large library of candidate terms that best represent the training data \citep{rudy_data-driven_2017}. Given time-series data of state variables $\boldsymbol{u}(x,t)$, the approach seeks a sparse coefficient vector $\boldsymbol{\xi}$ such that the time derivative is well approximated by a linear combination of candidate non-linear terms:
\begin{equation}
  \frac{\partial \boldsymbol{u}}{\partial t}
  \;\approx\;
  \boldsymbol{\Theta}(\boldsymbol{u},\,\nabla\boldsymbol{u},\,
  \nabla^2\boldsymbol{u},\,\ldots)\;\boldsymbol{\xi},
  \label{eq:sindy}
\end{equation}
where $\boldsymbol{\Theta}$ is a library matrix whose columns contain all candidate non-linear terms constructed from the state variables and their spatial derivatives up to a specified order. Because the derivatives appearing in $\boldsymbol{\Theta}$ are not always directly available from the data, they must often be approximated numerically, for example using finite-difference schemes. The accuracy of this approximation can significantly influence the convergence and stability of the sparse regression process. The choice of which terms to include in the library is similarly critical, as the library can grow rapidly for high-dimensional problems, increasing computational cost and the risk of spurious model selection.

The defining feature of sparse regression is the enforcement of sparsity, the optimisation seeks the fewest non-zero terms in $\boldsymbol{\xi}$ that adequately minimise the residual. This yields a Pareto front, trading model complexity against fitting error, and the preferred model is selected at the point where adding further terms produces only marginal improvement.

A widely used implementation of sparse regression for dynamical systems is the Sparse Identification of Non-linear Dynamics (SINDy) method \citep{brunton_discovering_2016}, together with its open-source implementation PySINDy \citep{kaptanoglu_pysindy_2022, brunton_promising_2024}. SINDy applies a Lasso regression, where sequential thresholded least-squares regression is applied to the formulation in Eq.~\eqref{eq:sindy}, iteratively removing terms with small coefficients until a parsimonious model is obtained. Related approaches include implicit SINDy \citep{kaheman_sindy-pi_2020}, which can handle implicit relationships between terms, and PDE-Net \citep{long_pde-net_2019}, which is notable for its ability to eliminate terms from the library during training, thereby reducing computational overhead. A systematic benchmarking of several equation discovery tools, including SINDy, has been performed by \citet{landajuela_unified_2022}, providing guidance on the relative strengths of different methods across a range of canonical test problems.

\begin{figure}
    \centering
    \includegraphics[width=0.9\linewidth]{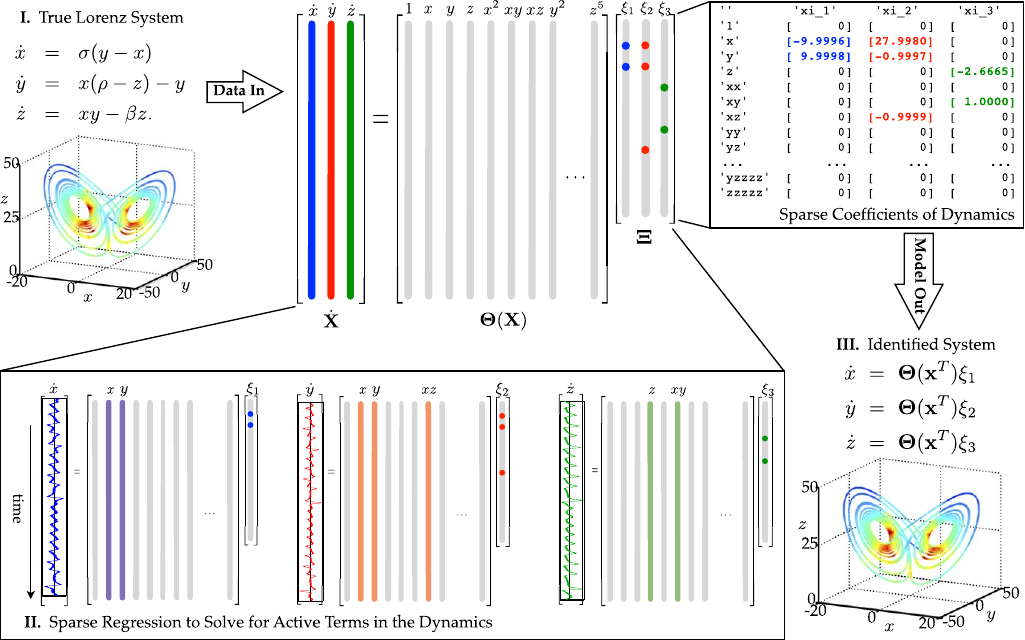}
    \caption{Schematic of the Sparse Identification of Non-linear Dynamics (SINDy) algorithm. Time-series data of state variables are collected (left), a library of candidate non-linear terms is constructed (centre), and sparsity-promoting regression identifies the minimal set of terms that best describe the dynamics (right), yielding an interpretable 
    governing equation. Reproduced from \citet{brunton_discovering_2016}, \textit{Proc.\ Natl.\ Acad.\ Sci.} \textbf{113}, 3932-3937, Fig.~1.}
    \label{fig:sindy_schem_fig}
\end{figure}

\begin{figure}
    \centering
    \includegraphics[width=0.9\linewidth]{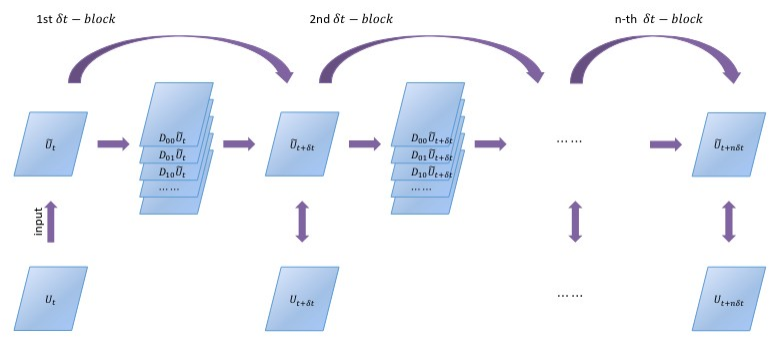}
    \caption{Schematic of the PDE-Net 2.0 architecture for data-driven PDE discovery. The network is constructed by stacking multiple $\delta t$-blocks into a feed-forward deep network, where all blocks share the same learnable parameters. Each $\delta t$-block takes the current state~$U_t$ as input and produces an updated state~$\tilde{U}_{t+\delta t}$ by combining two components: (i)~learnable convolution filters that approximate spatial differential operators (e.g.\ $\partial_x$, $\partial_{xx}$), with constraints on their moment matrices to ensure correspondence with finite-difference stencils; and (ii)~a symbolic neural network (SymNet) that approximates the non-linear response function~$F(U, \nabla U, \nabla^2 U, \ldots)$ of the underlying PDE. The forward Euler temporal discretisation makes the architecture ResNet-like, enabling long-term prediction by propagating the input through successive $\delta t$-blocks. During training, the accumulated prediction error $\|U(t + n\delta t) - \tilde{U}(t + n\delta t)\|^2$ is minimised jointly over all blocks, and sparsity-promoting regularisation on the SymNet parameters allows irrelevant terms to be driven to zero, thereby recovering an interpretable analytic form of the governing PDE. Reproduced from \citet{long_pde-net_2019}, \textit{J.\ Comput.\ Phys.}\ \textbf{399}, 108925, Fig.~2.}
    \label{fig:pdenet-schematic}
\end{figure}

Despite these strengths, sparse regression does require a degree of prior knowledge. The candidate library must be specified by the user, and for high-dimensional problems the number of terms can grow combinatorially, necessitating either physical insight to pre-select relevant terms or algorithmic strategies to prune the library. The method is also limited to discovering expressions within the span of the chosen library, and will fail to identify the correct model if the true functional form lies outside this space.

Another well-known challenge for all sparse regression methods is sensitivity to noise in the data, as pointwise derivative approximations amplify measurement errors. Several approaches mitigate this issue. The Weak SINDy (WSINDy) framework \citep{messenger_weak_2021, reinbold_using_2020} replaces pointwise derivative evaluations with integrals against smooth, compactly supported test functions, a formulation that naturally reduces noise. This approach is beneficial not only for data with additive measurement noise but also when the terms involve higher-order derivatives, where the choice of test-function kernel determines the degree of noise suppression. In the specific context of learning fluid equations from kinetic data, \citet{alves_data-driven_2022} demonstrated that formulating the regression problem in integral form significantly reduces systematic noise, an approach that can be viewed as a special case of the weak SINDy framework using box-kernel test functions. A related technique by \citet{stephany_pde-read_2022} combines sparse regression with weak PDE discovery, integrating the candidate equations over domains with smooth weight functions to remove derivatives from the response, and has been shown to handle substantial noise levels. An alternative strategy for noise robustness is to combine sparse regression with physics-informed neural networks: \citet{both_deepmod_2021} demonstrated that using PINNs to first learn a smooth representation of the solution field, from which derivatives can be accurately extracted, significantly improves SINDy performance in noisy settings. See also \citet{minor2025learning} for a related noise mitigation technique applied to terrestrial weather modelling.

It is important to distinguish sparse regression from the broader class of symbolic regression methods \citep{camps-valls_discovering_2023}. Sparse regression operates within a \emph{predefined} library of candidate terms and can therefore be regarded as a special case of symbolic regression over a restricted function space. General symbolic regression, by contrast, searches over a much larger space of mathematical expressions, defined by admissible functions and algebraic operations, without a fixed library. Because such spaces are combinatorially vast, symbolic regression typically relies on stochastic search strategies such as genetic algorithms to remain tractable \citep{cranmer_discovering_2020}. Model complexity can still be penalised in symbolic regression, but the absence of a fixed library means that the search is inherently less constrained. Both approaches produce interpretable, closed-form expressions; however, sparse regression is generally more computationally efficient when a suitable library can be constructed from prior physical knowledge, while symbolic regression offers greater flexibility when the functional form of the governing equations is largely unknown. Outside plasma physics, symbolic regression based on genetic programming has been applied to subgrid-scale closures in ocean and climate modelling, where it has been shown to produce compact, physically motivated expressions that can be tested both offline and online within numerical solvers \citep{ross_benchmarking_2023}. More recently, \citet{ying_neural_2025} combined symbolic regression with large language models to discover an expression for plasma-sheet pressure, illustrating the potential for hybrid AI-driven equation discovery in space-plasma contexts.

The principal advantage of equation discovery, whether via sparse or symbolic regression, is the interpretability of the resulting models, in a contrast to the 'black-box' nature of neural network surrogates. Because the discovered closure takes the form of an explicit mathematical expression, it can be directly analysed for physical consistency. For example, one can verify whether the expression satisfies expected symmetries of the underlying equations, such as rotational, translational, and Galilean (or, in the relativistic case, Lorentz) invariance. One can also assess thermodynamic consistency, for instance by checking that the closure does not generate negative entropy production. Such analyses are rarely performed in the literature but are important for ensuring that data-driven closures are physically admissible. Recent work by \citet{mcgrae-menge_embedding_2026} has shown that embedding symmetry constraints into the training process significantly improves both the accuracy and physical consistency of sparse regression closures for reconnection plasmas.

Interpretability also enables equation discovery to serve as a tool for informing theoretical work. By recovering a hierarchy of closure models of varying complexity, one can identify which physical approximations correspond to each level of truncation. An example is the work of \citet{alves_data-driven_2022}, who used sparse regression on PIC simulation data to obtain closures for a two-stream-unstable plasma, linking each discovered term to a known physical mechanism. This approach has been extended to multi-species modelling by \citet{ingelstenDatadrivenMultispeciesHeat2026}, who demonstrated that neural networks can also be a useful preliminary step in this process: by first confirming that a sufficiently accurate model exists within a given input space, neural networks provide an accuracy baseline against which the fidelity of the sparser regression model can be judged (see Section~\ref{sec:data_eq_disco}).

\section{Machine Learning Closures}

For the plasma closure problem, our surrogate model can either be an approximation for the entire system of plasma fluid equations, including a closure relation implicitly, or an approximation of the closure relation only, which is then integrated into the full plasma fluid model. Machine learning has additionally been applied to plasma modelling beyond the closure problem itself, for instance as a full surrogate that emulates the system dynamics directly rather than supplying a closure term: \citet{shekarpaz_surrogate_2025}, for example, trained a Deep Operator Network \citep{lu_deeponet_2020} to map initial plasma conditions to the time evolution of the electric-field energy under Landau damping, replacing the kinetic simulation entirely. Such full-surrogate approaches lie outside the scope of the present review, which is concerned specifically with closure relations for fluid models.

Before surveying individual studies, we introduce a terminological distinction that will be used throughout: a closure method is said to undergo \emph{offline} (\textit{a priori}) testing if the trained model is evaluated by comparing its closure outputs to a reference (analytic or kinetic) dataset without being embedded in a time-evolving fluid solver; \emph{online} (\textit{a posteriori}) testing refers to embedding the trained closure inside a fluid solver and evolving the system forward in time, so that errors in the closure accumulate dynamically. Online testing is substantially more demanding than offline testing and provides a direct measure of long-time stability and physical fidelity in a simulation context. As Table~\ref{tab:closure-summary} shows, online testing has been achieved by only a small minority of the studies reviewed here.

\subsection{Analytic Closure Training Data}
\label{sec:analytic_training}

Early work on constructing surrogate models for the plasma closure relation focused on proof-of-concept studies in which neural networks were trained on known analytic closures to establish whether even simple architectures could capture the relevant kinetic dynamics to a suitable degree of accuracy. These studies typically employed classical neural network architectures, such as the multi-layer perceptron and the convolutional neural network, and demonstrated that the neural network approach can reproduce the behaviour described by analytic closure models.

One of the earliest investigations into neural networks as surrogate closures came from \citet{ma_machine_2020}, who trained three architectures: a multi-layer perceptron (MLP), a convolutional neural network (CNN), and a two-layer discrete Fourier transform (DFT) network, on data generated from the Hammett-Perkins closure in configuration space:
\begin{equation}
  \tilde{q}(x) = -n_0 \sqrt{\frac{8}{\pi^3}}\, v_t
  \int_0^{\infty} \mathrm{d}x'\,
  \frac{\tilde{T}(x+x') - \tilde{T}(x-x')}{x'},
  \label{eq:hp-config}
\end{equation}
where the authors have performed an inverse Fourier transform on the Hammett-Perkins closure of equation~\eqref{hp-closure}. Throughout Sections~\ref{sec:analytic_training}-\ref{sec:kin_sim_training}, tildes denote first-order perturbation quantities about the equilibrium state, so that $\tilde{n} = n - n_0$, $\tilde{T} = T - T_0$, and $\tilde{q} = q - q_0$, where the subscript $0$ indicates the unperturbed background value. The configuration-space formulation was chosen specifically to probe non-local behaviour, which is absent from the Fourier-space form of the closure but arises through the dependence on temperature at neighbouring locations via the $\tilde{T}(x + x')$ terms.

The space was discretised into 128 grid points, with the temperature values at each grid point serving as input features and the corresponding heat flux values as targets. The authors identified a minimum volume of training data required for the deep learning methods to converge, and, more notably, found that the MLP required a minimum number of hidden-layer neurons corresponding to the number of Fourier degrees of freedom of the closure, despite training being conducted entirely in configuration space. The authors also investigated the effects of adding noise to the training data, finding that noise had no measurable effect when its magnitude was below the intrinsic prediction error of the network. Finally, the study highlighted key design considerations, including the locality of the target closure, the level of systemic noise in the data, and the scale and dimensionality of the problem, that influence the choice of architecture for future closure-learning efforts.

The Braginskii closure \citep{braginskii_transport_1965} is the classical transport description for collisional, short-mean-free-path plasmas: derived from a Chapman-Enskog expansion about a local Maxwellian, it gives strictly local relations in which heat flux and viscous stress follow from gradients of temperature and velocity with collision-dependent transport coefficients, in contrast to the non-local Landau-fluid closures above.

Building on this foundation, \citet{maulik_neural_2020} extended the neural network approach to analytic closure relations across multiple plasma regimes. The study considered three closures representing distinct physical assumptions: the Braginskii closure \citep{braginskii_transport_1965} for the collisional regime (fully local), the Hammett-Perkins closure \citep{hammett_fluid_1990} for the collisionless limit with periodic boundaries (globally non-local), and the Guo-Tang closure \citep{guo_parallel_2012} for the collisionless regime with an upstream Maxwellian source and a downstream absorbing wall (non-local over a finite region).

For the Braginskii closure, the lower-order moments (density, velocity, velocity gradients, and temperature) served as inputs, with the diagonal components of the viscosity tensor as targets. For the Hammett-Perkins closure, temperature values at grid points were mapped to heat flux values. For the Guo-Tang closure, profiles of the magnetic field and ambipolar potential were used as inputs and the three components of the heat flux as targets.

Three architectures were compared: a fully connected network, a locally connected network, and a convolutional network. The central finding was a direct link between the spatial locality of the target closure and the optimal degree of network connectivity: spatially local closures, such as the Braginskii model, were best captured by networks with restricted connectivity, whereas the globally non-local Hammett-Perkins closure required a fully connected architecture. This correspondence between closure locality and network connectivity provides guidance for selecting architectures in future data-driven closure work. It should be noted that this study was conducted entirely offline: the trained networks were evaluated against the analytic closure outputs but were not embedded into a running fluid solver. Integration of a trained neural-network closure into a plasma fluid simulation was first demonstrated by \citet{wang_deep_2020}, as discussed below.

The first online test of a neural-network closure came from \citet{wang_deep_2020}, who trained a multilayer perceptron on a kinetic Landau-fluid closure with collisions and periodic boundaries and embedded it in a fluid simulation.

The training dataset was generated from the Landau fluid closure model, which reduces to the Hammett-Perkins closure in the collisionless limit \citep{wang_landau-fluid_2019}:
\begin{equation}
  \tilde{q} = -\mathrm{i}\sqrt{2}\,\chi\, n_0 v_t\,
  \frac{k}{|k|}\,\tilde{T},
  \label{eq:landau-fluid}
\end{equation}
\begin{equation}
  \chi = \mathrm{i}\,
  \frac{(2\zeta^3 - 3\zeta)\,Z(\zeta) + 2\zeta^2 - 2}
       {(2\zeta^2 - 1)\,Z(\zeta) + 2\zeta},
  \label{eq:chi}
\end{equation}
where $\zeta = \omega/\sqrt{2}|k|v_t$ is the normalised phase speed and $Z(\zeta)$ is the plasma-dispersion function. $n_0$ is the unperturbed particle density and $v_t$ is the thermal speed. Input data was generated as perturbed temperature profiles, with the target heat flux calculated from equation~\eqref{eq:landau-fluid}. The authors trained MLP models on both collisional data in the static limit and on a collisionless dataset, both for cases of Landau damping from the closure model of \citep{wang_landau-fluid_2019}.

The neural network surrogate achieved higher accuracy than the non-Fourier numerical method for the closure of equation~\eqref{eq:landau-fluid}. When the trained model was embedded in a fluid simulation of single-mode Landau damping, the approximation errors introduced by the neural network remained smaller than the numerical errors produced by the non-Fourier method when both were propagated to late times. While this is an encouraging result, it was demonstrated only for one-dimensional Landau damping; the propagation of approximation errors in higher-dimensional and non-linear settings, where such errors may grow more rapidly, has yet to be examined.

This review is primarily concerned with the research conducted on closure relations for collisionless plasma, however, work done in the context of collisional plasma, has also been conducted, providing additional examples of ML techniques applied to plasma closure relations. \citet{lee_supervised_2023} applied supervised multivariate regression to learn parallel closure coefficients based off of the theoretical work from \citet{jiClosureTheoryHighcollisionality2023} (parallel heat flux and related parallel transport quantities) for a high-collisionality deuterium-carbon (DC) tokamak-edge plasma. Rather than training on raw kinetic data, they fit theoretical closure coefficients to produce compact analytical closure formulas suited to multi-ion, high-collisionality edge conditions. Beyond multi-ion transport, machine-learning closures have also been pursued for collisional kinetic models more broadly: \citet{christlieb_hyperbolic_2025} constructed hyperbolic machine-learning moment closures for the Bhatnagar-Gross-Krook (BGK) \citep{bhatnagarModelCollisionProcesses1954} equations, enforcing hyperbolicity of the resulting moment system as a stability constraint. In the related setting of non-local electron heat transport, \citet{luoLearningHeatTransport2025, luoResolutionIndependentMachineLearning2026, luoTimeembeddedConvolutionalNeural2026} trained a range of neural-network closures (theory-informed, time-embedded convolutional, and Fourier-operator architectures) on fully kinetic PIC data to learn heat-flux closures spanning the collisional-to-collisionless transition; this line of work was developed primarily for inertial-confinement and laser-plasma conditions but is methodologically relevant to the non-local closures discussed here.

\subsection{Kinetic Simulation Training Data}
\label{sec:kin_sim_training}

\subsubsection{Magnetic Reconnection and Pressure Tensor Closures}

Among the first to train neural networks on fully kinetic data were \citet{laperre_identification_2022}, who derived local closures for the electron pressure tensor and heat-flux vector and were one of few to target the full pressure tensor. Training data were extracted from fully kinetic simulations of a double-Harris sheet reconnection configuration, produced using the implicit Particle-in-Cell code iPiC3D \citep{markidis_multi-scale_2010}. Four simulations were generated with identical initial conditions except for the background guide field, which was varied between runs. Two machine learning models were trained: a multilayer perceptron (MLP) and a gradient-boosting regressor (GBR), with a baseline linear regressor included for comparison.

Training, validation and test sets were constructed by assigning different time snapshots from each simulation to each dataset, with cells selected preferentially from regions of high agyrotropy to focus on the reconnection dynamics. The input features were the magnetic field vector, its gradients and magnitude, the velocity vector, the charge density, and the parameter $\alpha = n_*^3/B_*^2$ from the closure of \citet{le_equations_2009}, whose parallel and perpendicular pressure components are given explicitly in Eqs.~\eqref{eq:le_ppar} and \eqref{eq:le_pperp}. The targets were the three components of the heat flux vector and the six independent components of the symmetric pressure tensor. To be clear, the machine learning models were not trained on data generated by the \citet{le_equations_2009} closure itself; the choice of input features was simply informed by the terms appearing in that closure relation.

Both machine learning models matched or exceeded the empirical closure and were consistently more accurate than the linear baseline. The MLP was more accurate than the GBR for the diagonal pressure tensor components across the reconnection region, while the GBR performed better away from the reconnection site. Both models struggled with the off-diagonal pressure tensor components and the heat flux vector, with the authors identifying the sampling strategy, limited data diversity, and PIC particle noise as contributing factors. The off-diagonal components of the electron pressure tensor emerged as a particular challenge, a difficulty that recurs in later studies using different methods (see Section~\ref{sec:data_eq_disco}).

The study is also among the first to use two-dimensional PIC data, in contrast to the predominantly one-dimensional problems considered in earlier work, providing early insight into the additional challenges that arise when extending data-driven closure methods to higher dimensions.

The neural-network approach was carried into the more realistic setting of the turbulent magnetosheath by \citet{miloshevich_electron_2025}. The authors trained a fully convolutional neural network (FCNN) to learn a non-local electron pressure tensor closure for a five-moment fluid model. The convolutional architecture enforces translational invariance and captures non-local spatial correlations in the pressure tensor without requiring the full simulation domain as input.

Training data were drawn from PIC simulations with 256 particles per cell, while the test set used higher-fidelity simulations with 5000 particles per cell at slightly different plasma parameters, providing a stringent test of generalisation across both simulation fidelity and physical parameter space. The FCNN was substantially more accurate than a baseline MLP and than both the CGL and empirical \citet{le_equations_2009} closures in all accuracy metrics, including the pressure-strain interaction and agyrotropy. A noteworthy finding was that a variant omitting the electric field from the inputs (FCNNnoE) showed the most robust generalisation to the high-fidelity test set, suggesting that including the electric field introduces spurious correlations during training.

This study, alongside the earlier work of \citet{wei_data-driven_2023}, is one of the few to test whether neural network closures can generalise across different simulation datasets. Online testing of the FCNN closure within a running fluid solver has not yet been attempted.

\subsubsection{Landau Damping Closures}

In one of the first applications of PINNs to the closure problem, \citet{qin_data-driven_2023} learned an implicit closure from one-dimensional Landau-damping data using both PINNs and gradient-enhanced PINNs (gPINNs).

\begin{figure}
    \centering
    \includegraphics[width=0.9\linewidth]{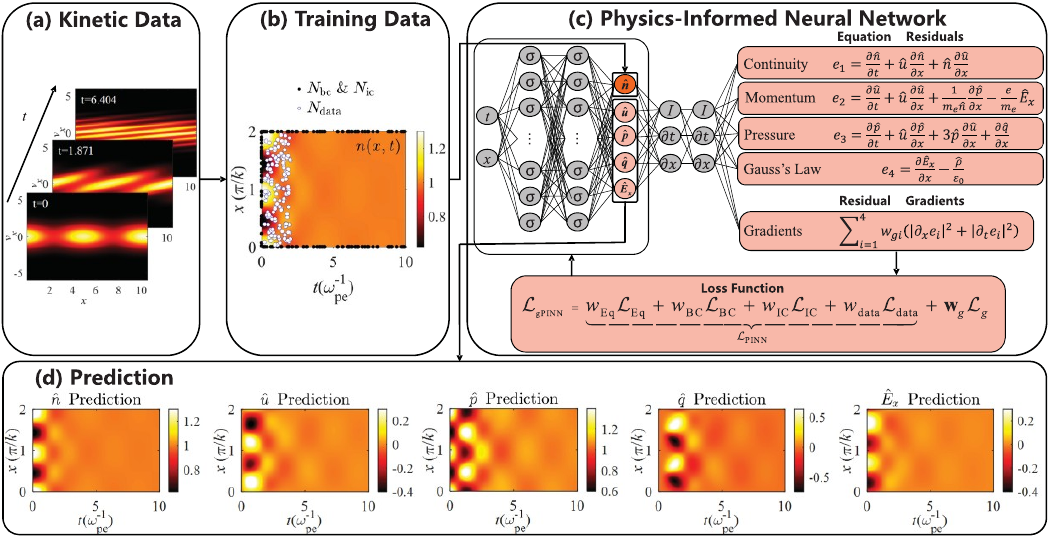}
    \caption{Overview of the physics-informed neural network (PINN) methodology for learning an implicit plasma closure from one-dimensional Landau-damping kinetic simulation data. (a)~Spatiotemporal evolution of the kinetic simulation data, showing the density, bulk velocity, pressure, heat flux, and electric field. (b)~Subdomains from which the training data are sparsely sampled, restricted to early simulation times to test the model's ability to extrapolate. (c)~Architecture of the PINN: the network takes time and spatial coordinates as inputs and outputs all fluid variables simultaneously, with the loss function incorporating the fluid moment equations as physics-informed residual constraints alongside the data loss. The gPINNp variant additionally penalises the gradient of the pressure residual only, exploiting the expected sensitivity of the pressure evolution to the closure. (d)~Model predictions for each fluid variable, demonstrating good agreement with the kinetic simulation even at late times beyond the training window. The closure is learned implicitly within the coupled system of equations rather than as a standalone module. Reproduced from \citet{qin_data-driven_2023}, \textit{Phys.\ Rev.\ Research}\\
  \textbf{5}, 033079, Fig.~1.}
    \label{fig:qin-pinn-fig}
\end{figure}

The authors introduced a new variant, the gPINNp, which penalises only the gradient of the pressure equation residual rather than the gradients of all residuals, substantially reducing the computational cost relative to the full gPINN. Formally, the standard PINN loss is
\begin{equation}
  \mathcal{L}_{\mathrm{PINN}}
  = \mathcal{L}_{\mathrm{data}}
  + \sum_i \lambda_i \,\|\mathcal{R}_i\|^2,
  \label{eq:pinn-loss}
\end{equation}
where $\mathcal{R}_i$ are the residuals of the $i$-th fluid equation evaluated at the collocation points. The gPINN augments this with gradient terms:
\begin{equation}
  \mathcal{L}_{\mathrm{gPINN}}
  = \mathcal{L}_{\mathrm{PINN}}
  + \sum_i \mu_i \left\|
      \frac{\partial \mathcal{R}_i}{\partial x}
    \right\|^2,
  \label{eq:gpinn-loss}
\end{equation}
and the gPINNp variant retains only the gradient of the pressure equation residual:
\begin{equation}
  \mathcal{L}_{\mathrm{gPINNp}}
  = \mathcal{L}_{\mathrm{PINN}}
  + \mu_p \left\|
      \frac{\partial \mathcal{R}_p}{\partial x}
    \right\|^2.
  \label{eq:gpinnp-loss}
\end{equation}

All three models predicted the plasma moments well into late times, with the gPINNp achieving the highest accuracy, particularly at later times. This result is physically consistent: the heat flux is most closely correlated with the pressure and its gradients, as expected for a Landau fluid closure, and the gPINNp exploits this sensitivity by directing the optimiser towards closure-relevant information without imposing additional constraints on the other moment equations.

As shown in Figure~\ref{fig:qin-pinn-fig}, the network takes time and spatial coordinates as inputs and outputs all fluid variables simultaneously, learning the entire system of fluid equations with an implicit closure. A practical limitation of this approach is that, because the network learns the coupled system jointly, the closure cannot be extracted as a standalone module and inserted into an independent fluid solver; the PINN therefore acts as a surrogate for the coupled fluid system rather than as a modular closure component.

The kinetic simulation was sampled sparsely at early times only, forcing the models to extrapolate to later times. The gPINNp model agreed well with the kinetic data at late times despite the absence of late-time training data, demonstrating the capacity of the physics-informed approach to extrapolate beyond the training distribution, a capability that standard neural networks typically lack.

Neural operators were first applied to the plasma closure problem by \citet{wei_data-driven_2023}, who used the Fourier Neural Operator (FNO) to learn a multi-moment fluid model from several sets of initial conditions.

The Fourier Neural Operator \citep[FNO;][]{li_fourier_2021} approximates solution operators of PDEs by composing a sequence of spectral convolution layers, each of the form
\begin{equation}
  v_{\ell+1}(x) = \sigma\!\left(
    \mathsfbi{W}_\ell \, v_\ell(x)
    + \mathcal{K}_\ell\bigl(v_\ell\bigr)(x)
    + \boldsymbol{b}_\ell
  \right),
  \label{eq:fno-layer}
\end{equation}
where $v_\ell : \mathbb{R}^d \to \mathbb{R}^{d_v}$ is the hidden representation at layer $\ell$, $\mathsfbi{W}_\ell$ is a local linear transform, $\boldsymbol{b}_\ell$ a bias, and $\sigma$ a pointwise non-linearity. The integral kernel operator $\mathcal{K}_\ell$ is implemented via the Fast Fourier Transform (FFT):
\begin{equation}
  \mathcal{K}_\ell(v)(x)
  = \mathcal{F}^{-1}\!\left(
      \mathsfbi{R}_\ell \cdot \mathcal{F}(v)
    \right)(x),
  \label{eq:fno-kernel}
\end{equation}
where $\mathcal{F}$ and $\mathcal{F}^{-1}$ denote the forward and inverse FFTs, and $\mathsfbi{R}_\ell$ is a learnable complex-valued weight tensor applied to the first $k_{\max}$ Fourier modes; all modes with wavenumber $|k| > k_{\max}$ are truncated to zero. This truncation enforces a low-frequency inductive bias that is well-matched to smooth, wave-like physical fields. The full network follows the structure outlined by equation~\eqref{eq:neuralop} from \citet{kovachki_neural_2024, li_fourier_2021}.

\begin{figure}
  \centering
  \includegraphics[width=0.9\textwidth]{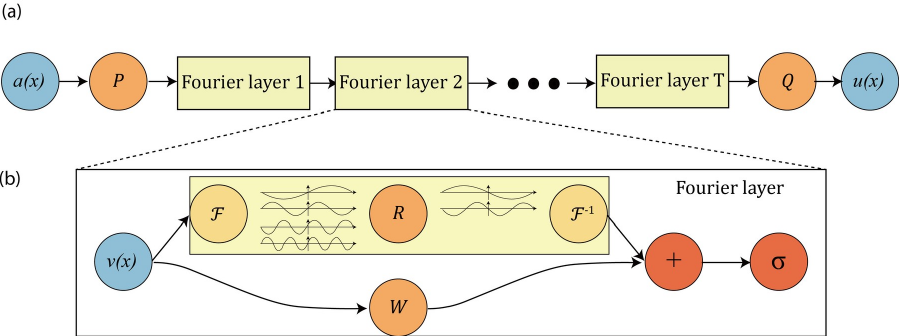}
  \caption{Architecture of the Fourier Neural Operator (FNO). 
    \textit{Top}: the full network pipeline from input function through 
    lifting~($P$), iterative Fourier layers, and projection~($Q$) to the 
    output function. 
    \textit{Bottom}: detail of a single Fourier layer, showing the 
    parallel paths of the spectral convolution 
    ($\mathcal{F} \to R_\ell \to \mathcal{F}^{-1}$, where high-frequency 
    modes above~$k_\mathrm{max}$ are truncated) and the local linear 
    transform~$W_\ell$, combined through a pointwise non-linearity~$\sigma$. 
    Reproduced from \citet{kovachki_neural_2024}, \textit{J.\ Mach.\ Learn.\ Res.}\ 
    \textbf{24}(89), 1-105, Fig.~5.}
  \label{fig:fno_architecture}
\end{figure}

The FNO achieved a 10.8\% relative error on the test set, compared with 66.7\% for the MLP baseline trained on the same data, highlighting the neural operator's advantage in learning from multiple sets of initial conditions simultaneously. Whether this superior performance reflects the intrinsic suitability of Fourier representations for wave-like closures or arises principally from the greater expressive capacity and efficient GPU implementation of the FNO architecture is not yet clear. The FNO's spectral basis may also be inherently well-matched to the Hammett-Perkins closure, which is local in Fourier space; \citet{ma_machine_2020} observed a similar advantage when using a discrete Fourier transform network. Disentangling the operator-learning framework from the Fourier representation will require application of neural operators to closures that are not naturally expressed in Fourier space.

A key practical advantage of the FNO is that the FFT maps efficiently to GPU hardware, so the non-local spectral kernel can be evaluated at $\mathcal{O}(N \log N)$ cost rather than $\mathcal{O}(N^2)$ for a dense integral kernel. This is directly relevant to the closure problem: the computational expense of non-local analytic closures such as Hammett-Perkins arises from the need to evaluate a global integral (or equivalently a Hilbert transform) at every grid point, and the Hilbert transform is intrinsically difficult to distribute across the domain decomposition used in parallel plasma codes. The FNO sidesteps this difficulty by implementing its non-local kernel entirely in Fourier space via the FFT (equation~\eqref{eq:fno-kernel}), making the evaluation of non-local closures competitive in cost with local analytic closures and addressing one of the longstanding practical objections to Hammett-Perkins-type heat fluxes in large-scale fluid codes \citep{li_fourier_2021}.

Closely related is the work of \citet{huang_machine-learning_2025}, who used the FNO to learn a heat-flux closure for both linear and non-linear Landau damping and, crucially, tested it online. Training data were generated from a Vlasov solver, with the Hammett-Perkins closure motivating the choice of a Fourier-based architecture. They then integrated the trained FNO closure into a fluid solver and evolved the system forward in time, achieving one of the few published examples of online closure testing in the ML plasma literature (see Table~\ref{tab:closure-summary}). The FNO-based fluid model reproduced both linear and non-linear Landau damping more accurately than the Hammett-Perkins closure, indicating that the data-driven model captured kinetic physics beyond the reach of the analytic closure.

A fundamentally different route to online closure was taken by \citet{joglekar_machine_2023}, who trained a compact network end-to-end through the dynamics of a running fluid solver using automatic differentiation. Their ADEPT framework augments the one-dimensional Euler equations with a hidden variable $\delta$ governed by a prognostic equation whose source term is parameterised by a compact feedforward network (4 layers, 8 neurons, 160 parameters, $\tanh$ activation). The hidden variable encodes the prior kinetic state, specifically the population of resonantly trapped electrons responsible for suppressing non-linear Landau damping, without resolving velocity space explicitly.

The network was trained end-to-end by differentiating through all timesteps of the fluid simulation and minimising the discrepancy with ground-truth Vlasov-Fokker-Planck simulations across 200 training runs varying collision frequency, perturbation amplitude, and wavenumber. The resulting model reproduced both the linear damping regime and the strongly non-linear bounce/saturation regime over five orders of magnitude in electric field energy. The model generalised to domains $20\times$ larger than the training domain with open (rather than periodic) boundary conditions, capturing the spatial non-locality of trapped-electron escape, and ran approximately $14\times$ faster than the kinetic solver. This study provided one of the clearest demonstrations that online training, where the network is optimised through its effect on the full simulation trajectory rather than by matching pointwise closure outputs, can yield generalisable and physically faithful closures even with an extremely compact network.

\subsubsection{General Fluid Closures}

\citet{bois_neural_2020} trained a fully convolutional V-net architecture to learn a heat flux closure for the one-dimensional Euler-Poisson system, using training data from Vlasov-Poisson kinetic simulations spanning a wide range of Knudsen numbers from the highly collisional to the collisionless regime. The V-net uses overlapping spatial windows to capture non-local interactions while maintaining computational scalability, and a per-Knudsen-number normalisation ensures uniform accuracy across the full range of collisionality. The study demonstrated that a single convolutional network, with an appropriate windowing strategy, can handle the transition from local to non-local closure behaviour within a unified model, spanning regimes where neither the Braginskii nor the Hammett-Perkins analytic closures are individually accurate. Testing was performed offline only.

A different methodology entirely was used by \citet{shukla_learned_2022}. Rather than training a neural network, the authors applied singular value decomposition (SVD) to gyrokinetic simulation data to extract the dominant kinetic modes driving phase mixing, constructing a `Learned Multi-mode' (LMM) closure in which the fluid model is projected onto the reduced SVD basis. The LMM closure was trained at only 3 points in a 35-point parameter space spanning temperature gradient and collision frequency. Despite this limited training, the closure achieved approximately 8\% root-mean-square error across the full parameter space (using three training points) and reproduced the reduction in phase-mixing rate observed in the non-linear turbulent regime relative to linear theory.

Crucially, the LMM closure was validated \textit{a posteriori} (online) rather than merely \textit{a priori} (offline). The closure coefficients, precomputed per wave vector from the singular value decomposition of the kinetic training run, were embedded directly within the four-moment fluid solver and used to supply the truncated moment at every timestep of a fully non-linear fluid simulation. Closure performance was then assessed from the resulting closed-loop dynamics compared against the kinetic reference. The reported root-mean-square errors of 8\% (three training points) and 12\% (two training points) in the saturated heat flux are therefore online errors accumulated over the full turbulent evolution, a considerably more stringent test than the \textit{a priori} comparisons that characterised many contemporaneous neural-network closures.

This online testing also exposed a failure mode invisible to \textit{a priori}
evaluation. When applied far from its training point, the closure trained at low
gradient drive and collisionality produced a fluid simulation that initially
saturated at the correct transport level before jumping spuriously to a higher
level partway through the run, generating large errors at the highest gradient
drives. Only by integrating the closure within the evolving system, rather than scoring isolated predictions, could this sensitivity to the dynamical trajectory be detected, underscoring both the value of online validation and the extrapolation risk inherent to sparsely trained data-driven closures.

A distinct strategy is to close the moment hierarchy in velocity space rather than configuration space. \citet{barbour_machine-learning_2025} used reservoir computing, a recurrent machine-learning architecture, to build such a closure for the one-dimensional collisionless Vlasov-Poisson system, treated as the electrostatic limit of gyrokinetics. Solving the system pseudo-spectrally on a Fourier basis in configuration space and a Hermite basis in velocity space recasts the Vlasov equation as an infinitely coupled hierarchy of fluid moments; exploiting the locality of interactions in the Hermite representation, the network modelled the small-scale velocity-space dynamics and closed the hierarchy at a chosen Hermite moment. Unlike the configuration-space and heat-flux closures discussed above, this closure acts directly on the velocity-space moment hierarchy. It was tested online within the running Fourier-Hermite solver: in the linear limit it reproduced the closed Hermite moment to within $2\%$ while reducing the velocity resolution, and in the strongly non-linear regime it resolved the low-order Fourier and Hermite spectra more accurately than a naive truncation closure, cutting the required velocity resolution by a factor of sixteen. The study is one of the few to close in velocity space and to target the electrostatic-gyrokinetic setting relevant to magnetised-turbulence applications.

\subsection{Data Driven Equation Discovery}
\label{sec:data_eq_disco}

In addition to the research into data-driven surrogates of the plasma closure, there are investigations into finding explicit analytic closure models directly from kinetic simulation data. Equation discovery encompasses many different machine learning methods for extracting models from data; the studies discussed here predominantly employ sparse regression, with one study by \citet{liu_data-driven_2022} using physics-informed neural networks in an inverse problem to recover the governing equations from a kinetic simulation.

In contrast to the surrogate model approach, equation discovery produces directly interpretable models that can inform theoretical development of further closure relations. The analytic expressions produced by equation discovery methods are also more straightforward to implement in fluid solvers than neural network surrogates, although they remain subject to the approximation errors inherent to all data-driven approaches, which must be quantified when integrating the discovered closures into fluid simulations.

\subsubsection{SINDy Approach}

\citet{alves_data-driven_2022} applied sparse regression to fully kinetic PIC simulation data to discover reduced models of plasma dynamics, seeking interpretable and generalisable expressions from which connections to existing theoretical work could be drawn.

A central challenge identified by the authors is the sensitivity of sparse regression to noise, which can be substantial in PIC simulations due to the projection of macro-particles onto the grid. To address this, the study reformulated the sparse regression problem to identify the governing PDEs in integral form, substantially reducing the effect of noise compared to the differential formulation. This integral approach can be viewed as a special case of the Weak SINDy (WSINDy) framework \citep{messenger_weak_2021, reinbold_using_2020}, which replaces pointwise derivative evaluations with integrals against smooth, compactly supported test functions.

\begin{figure}
    \centering
    \includegraphics[width=0.9\linewidth]{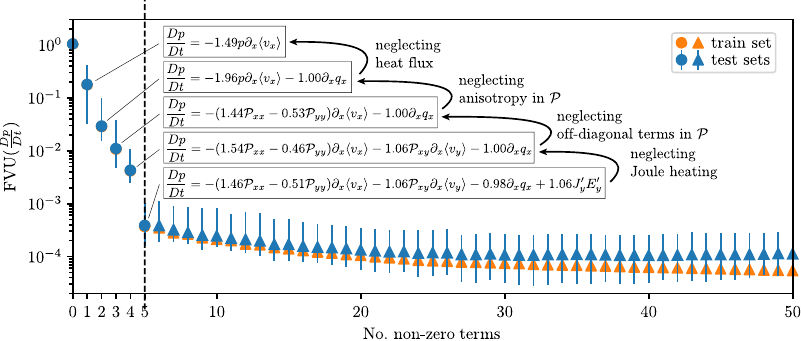}
    \caption{Accuracy-complexity Pareto front for closure models discovered via sparse regression (SINDy) applied to fully kinetic particle-in-cell simulation data. The horizontal axis gives the model complexity, measured by the number of non-zero terms retained in the sparse regression, while the vertical axis shows the fraction of variance unexplained (FVU). Orange circles denote training-set error and blue triangles denote test-set error, evaluated over a 30-fold cross-validation procedure. At each level of complexity, the spread of the markers reflects the variability of the discovered model across different cross-validation realisations; models whose terms are consistently identified across folds (low spread) are more robust than those exhibiting high variability between realisations. Annotations along the Pareto front identify the physical approximations corresponding to each level of complexity: from the simplest model (neglecting heat flux) through progressive inclusion of pressure anisotropy, off-diagonal pressure tensor terms, and Joule heating. The sharp knee at five non-zero terms marks the optimal model, where the Pareto front breaks clearly and the discovered terms are consistent across cross-validation folds; beyond this point, additional terms yield diminishing improvements in accuracy and the model form becomes less stable. This hierarchy demonstrates how sparse regression can recover and rank well-known closure approximations directly from kinetic data, providing a data-driven framework for guiding the development of new closure models. Reproduced from \citet{alves_data-driven_2022}, \textit{Phys.\ Rev.\ Research} \textbf{4}, 033192.}
    \label{fig:Alves_paper}
\end{figure}

Using this integral formulation, the authors recovered the coefficients of the kinetic Vlasov equation to within 2\% error, the coefficients of the multi-fluid equations to within 1-2\% error, and the coefficients of the one-fluid MHD equations to within 1-4\% error, demonstrating that sparse regression can robustly identify the governing equations of a plasma system from noisy kinetic data. Analysis of the accuracy-complexity trade-off (Figure~\ref{fig:Alves_paper}) reveals a Pareto front of reduced models with increasing complexity, where the preferred model is selected at the knee of the front, the point beyond which additional terms yield only marginal improvement. Each level of complexity in this hierarchy corresponds to a different closure approximation, from the simplest model (neglecting heat flux) through progressive inclusion of pressure anisotropy, off-diagonal pressure tensor terms, and Joule heating. The hierarchy can therefore be read as a data-driven ranking of the physical approximations underlying different closure models, providing a framework for guiding the development of new closures.

The integral formulation introduced here proved particularly influential: many subsequent sparse regression studies adopted the same approach to mitigate PIC noise. It should be noted, however, that the integral technique used by \citet{alves_data-driven_2022} is one instance of a broader class of weak-form methods; the WSINDy framework of \citet{messenger_weak_2021} and the approach of \citet{reinbold_using_2020}, offer more general formulations using smooth kernel functions rather than box integrals. Future work on improving closure relations will focus on utilising the hierarchy of models as a guiding framework for selecting the terms most important to the dynamics when constructing new closures. Throughout this work, the authors utilised SINDy primarily as a means of confirming the equations of motion of the system and then inferring the closure. This highlighted a limitation of the SINDy method when applied to the closure problem, as it is more suited to finding governing PDEs of the system, as opposed to explicit closure relations.

The sparse-regression approach was extended to a ten-moment reconnection model by \citet{donaghy_search_2023}, who used SINDy to discover a closure truncating the hierarchy at the pressure tensor. Training data were generated from fully kinetic PIC simulations of a collisionless Harris sheet reconnection problem using the PSC code \citep{germaschewski_plasma_2016}, with the integral formulation of \citet{alves_data-driven_2022} employed to mitigate particle noise.

The study first verified the ten-moment model against the PIC data, satisfying the zeroth-, first-, and second-order moment equations with normalised $L_2$ error $\leq 0.28$ for the on-diagonal terms. The off-diagonal terms, however, proved difficult to recover accurately, with the discovered equations exhibiting large coefficient errors, echoing the difficulties encountered by \citet{laperre_identification_2022} with the off-diagonal pressure tensor in a neural network context. The discovered solutions deviated narrowly from conservation laws, with SINDy eliminating low-magnitude terms that fell below the sparsity threshold; the authors argued this reflected the choice of elimination bounds rather than a genuine violation of conservation, and suggested that feature scaling could alleviate this issue.

When comparing the discovered closure to the local Hammett-Perkins baseline (most readily assessed via Table~5 of \citet{donaghy_search_2023}, which compares the tensor components directly), the on-diagonal terms showed significantly reduced numerical error, while the off-diagonal terms failed to improve over the baseline. The discovered closure performed well in the non-linear phase but could not be validated in the linear regime, where the signal-to-noise ratio was too low for reliable training. The persistent difficulty with the off-diagonal pressure components across both neural network \citep{laperre_identification_2022} and sparse regression approaches suggests that this challenge may be intrinsic to the reconnection geometry or the available data, rather than specific to any one method. Whether three-dimensional data, larger training sets, or alternative architectures can resolve this is not yet known.

\citet{ingelsten_data-driven_2024} applied sparse regression to one-dimensional PIC simulations of Landau damping and the two-stream instability to discover a heat flux closure and quantify the relative importance of its constituent terms. Model complexity was controlled by varying the target number of retained terms, with the integral technique utilised by \citet{alves_data-driven_2022} again used to suppress simulation noise.

For both Landau damping and the two-stream instability, the sparse regression consistently identified a six-term optimal closure. By selectively removing individual terms and measuring the resulting change in approximation error, the authors quantified the relative contribution of each term. One term was found to be disproportionately important during the growth/decay phase, suggesting it encodes the physics of wave-particle energy exchange, while other terms contributed primarily in the saturated state. This term-by-term analysis illustrates how sparse regression can be used not only to discover closures but also to diagnose which physical processes dominate in different dynamical regimes.

The study did not include terms coupled to the $E$ and $B$ fields in the candidate library, which would greatly increase the library size and computational cost. The closure was also restricted to a local approximation; a non-local extension has not yet been attempted.

\citet{ingelstenDatadrivenMultispeciesHeat2026} later generalised this six-term closure from a single combined electron species to multiple electron species, a description better suited to fluid modelling of two-stream-unstable plasmas. Training data were generated from OSIRIS particle-in-cell simulations of an electron-proton plasma over a range of two-stream initial conditions. Because the discovered closure coefficients were found to vary with plasma conditions, the authors developed a non-linear sparse regression (NLSR) framework that fits rational-function models for the three most important coefficients in terms of box-averaged fluid quantities. Neural networks were trained on the same inputs to establish an accuracy baseline, providing an approximate lower bound on the achievable error against which the sparser NLSR models could be judged; the NLSR models reached this neural-network-equivalent accuracy while remaining substantially more interpretable and using over an order of magnitude fewer parameters.

These closures capture $80$--$90\%$ of the variation in the species heat fluxes and predict the pressure evolution $\partial_t p_\sigma$ with a typical accuracy of $85$-$95\%$, compared with roughly $50$-$60\%$ for a naive zero-heat-flux closure. As with the single-species study, testing was performed offline.

A common limitation of the equation discovery studies discussed above is that the discovered closures are not guaranteed to respect the fundamental symmetries of the underlying kinetic equations. \citet{mcgrae-menge_embedding_2026} addressed this directly by embedding physical symmetries into the training process via data augmentation. Applied to two-dimensional PIC simulations of magnetic reconnection, the approach augmented the training data with Galilean and Lorentz boosts to generate additional samples that are physically equivalent but expressed in different reference frames.

The symmetry-augmented training was applied to both sparse regression and neural network models for the electron pressure tensor closure. The augmented models clearly surpassed both their un-augmented counterparts and commonly used theoretical closures, including CGL and \citet{le_equations_2009}. Augmentation also improved data efficiency: models trained with symmetry-augmented data required fewer training samples to reach a given accuracy, an important consideration given the computational expense of generating PIC training datasets. This approach provides a general strategy for incorporating known physical constraints into data-driven closure models without modifying the underlying machine learning architecture, and could be combined with any of the sparse regression or neural network methods discussed in this review.

\subsubsection{Neural Network Approach}

\citet{liu_data-driven_2022} approached equation discovery through a PINN, recovering both the ten-moment fluid equations and a heat-flux closure from one-dimensional Landau-damping data. The authors employed two PINNs: one learning the fluid model coefficients from the kinetic data, and the other forming a surrogate model trained to predict the output. The method yielded the following closure relation for the heat flux:
\begin{equation} \label{liu_closure}
    q = -4.1231 \sqrt{\frac{p}{n}} \left( \frac{\partial p}{\partial x} - \frac{p}{n} \frac{\partial n}{\partial x} \right).
\end{equation}
The coefficients recovered by the PINN lie within 1\% of the theoretical 10-moment fluid model values, and the closure predictions agree well with the kinetic simulation data for both density and electric field. However, the work is confined to one dimension and to fitting coefficients of a pre-defined closure form, so the technique does not yet discover genuinely new functional forms. The study demonstrated that a PINN-based inverse problem approach is viable and comparable in performance to sparse regression for extracting governing equations from kinetic data.

\citet{cheng_data-driven_2023} pursued equation discovery with PDE-net \citep{long_pde-net_2019}, a method that combines sparse regression with neural networks and can prune library terms during training. Applied to spatio-temporal data from a one-dimensional kinetic simulation of Landau damping, PDE-net recovered the multi-moment fluid model together with the following heat flux closure:
\begin{equation}
    q = -4.2665 \, n \sqrt{T} \frac{\partial T}{\partial x},
\end{equation}
which, under the substitution $T = p/n$, is identical in functional form to the closure discovered by \citet{liu_data-driven_2022} (equation~\ref{liu_closure}), with only a modest difference in the numerical coefficient. The learned model correctly reproduced the Landau damping rates observed in the kinetic simulation.

A distinguishing feature of this study is the use of time-prediction as a validation step: the discovered multi-moment model (with closure) was propagated forward in time using forward Euler, and the predicted fluid quantities were compared with the kinetic data. Agreement was good within the investigated time interval, although errors grew at late times, likely due to small inaccuracies in the learned coefficients compounding through the time-stepping scheme. The PDE-net method, while leveraging the predictive capabilities of neural networks, still requires expected terms to be included in the candidate library, and the study considered only simple one-dimensional Landau damping with a high level of prior knowledge of the expected model. Extension to higher dimensions and more complex phenomena with unknown closure forms is still unresolved, as is the source of the late-time error growth that would need to be controlled before integration into large-scale simulations.

\subsubsection{Summary of Equation Discovery Progress}

Sparse regression has been the most widely used equation discovery technique for plasma closures, but PDE-net and PINNs have also demonstrated viability. Most studies to date have been proof-of-principle demonstrations confined to a single spatial dimension; only \citet{donaghy_search_2023} has applied sparse regression to a two-dimensional problem. A common theme across these studies is the capacity of equation discovery to extract physical insight from the data: the Pareto hierarchy of \citet{alves_data-driven_2022} ranks closure approximations by their contribution to model accuracy, while the term-removal analysis of \citet{ingelsten_data-driven_2024} quantifies the importance of individual terms in different dynamical regimes. This interpretability is a distinctive advantage of equation discovery over the neural network surrogate approach.

Several challenges remain. Integration of discovered closures into time-evolving fluid solvers has received limited attention, with only preliminary discussion from \citet{ingelsten_data-driven_2024, ingelstenDatadrivenMultispeciesHeat2026} and \citet{cheng_data-driven_2023}; the numerical stability of these models in high-dimensional, long-time simulations is largely untested. The candidate term libraries used by sparse regression grow combinatorially with dimensionality and non-linearity, and strategies for managing this growth, whether through physically motivated library pruning, hierarchical search, or the use of general symbolic regression methods that do not require a fixed library, will be essential for extending equation discovery to more complex plasma dynamics.

\subsection{Summary of Machine Learning Closure Studies}

Table~\ref{tab:closure-summary} summarises all machine learning plasma closure studies discussed in this section, organised by subsection and indicating for each whether testing was performed offline or online.

{\small
\begin{longtable}{p{1.6cm} p{2.0cm} p{2.2cm} p{2.3cm} p{1.0cm} p{2.8cm}}
\caption{Summary of machine learning plasma closure studies discussed in Section 3. ``Online'' testing denotes embedding the trained model inside a running fluid solver; ``Offline'' denotes validation against a kinetic or analytic reference dataset without time-stepping.}
\label{tab:closure-summary}\\
\toprule
\textbf{Reference} &
\textbf{ML Method} &
\textbf{Training Data} &
\textbf{Target Closure} &
\textbf{Online / Offline} &
\textbf{Key Finding} \\
\midrule
\endfirsthead
\multicolumn{6}{l}{\textit{Table~\ref{tab:closure-summary} continued}}\\
\toprule
\textbf{Reference} &
\textbf{ML Method} &
\textbf{Training Data} &
\textbf{Target Closure} &
\textbf{Online / Offline} &
\textbf{Key Finding} \\
\midrule
\endhead
\bottomrule
\endfoot

\multicolumn{6}{l}{\textit{Section 3.1: Analytic closure training data}}\\
\midrule

\citet{ma_machine_2020}
  & MLP, CNN, DFT-net
  & Analytic (Hammett-Perkins formula)
  & Heat flux (config.\ space)
  & Offline
  & Minimum training volume required; MLP neuron width linked to Fourier degrees of freedom. \\

\citet{maulik_neural_2020}
  & MLP, CNN, locally connected net
  & Analytic (Braginskii, HP, Guo-Tang formulae)
  & Heat flux / viscosity (three closure types)
  & Offline
  & Architecture must match closure locality; fully connected net needed for non-local closures. \\

\citet{wang_deep_2020}
  & MLP
  & Analytic (Landau fluid / non-Fourier closure)
  & Heat flux (Landau fluid)
  & \textbf{Online}
  & First published NN closure integrated into plasma fluid solver; NN errors smaller than numerical errors at late times. \\

\citet{lee_supervised_2023}
  & Supervised ML regression
  & Theoretical closure coefficients
  & Parallel closures, deuterium-carbon tokamak-edge plasma
  & Offline
  & Compact analytical closure formulas for multi-ion high-collisionality plasma. \\

\midrule
%%% --- Section 3.2: Kinetic simulation training data ---
\multicolumn{6}{l}{\textit{Section 3.2: Kinetic simulation training data}}\\
\midrule

\citet{laperre_identification_2022}
  & MLP, gradient-boosting regressor
  & PIC (iPiC3D), magnetic reconnection
  & Pressure tensor + heat flux
  & Offline
  & ML matches/exceeds \citet{le_equations_2009} closure for diagonal $\mathsf{P}$; off-diagonal hardest. \\

\citet{miloshevich_electron_2025}
  & FCNN (fully convolutional)
  & PIC (magnetosheath turbulence)
  & Electron pressure tensor, 5-moment model
  & Offline
  & Outperforms MLP, CGL, \citet{le_equations_2009}; FCNNnoE most robust; generalises across fidelity levels. \\

\citet{bois_neural_2020}
  & Fully conv.\ V-net
  & Vlasov-Poisson simulations
  & Heat flux (Euler-Poisson), wide Knudsen range
  & Offline
  & Overlapping-window CNN handles non-locality; accurate across $\mathrm{Kn}\in[0.01,1]$. \\

\citet{shukla_learned_2022}
  & SVD basis (LMM)
  & Gyrokinetic simulations
  & Phase-mixing closure, gyrokinetic turbulence
  & \textbf{Online}
  & sparsely trained; tested online revealed unseen failure modes from offline testing; outperforms Hammett-Perkins. \\

\citet{barbour_machine-learning_2025}
  & Reservoir computing
  & Fourier-Hermite Vlasov-Poisson solver
  & Velocity-space (Hermite) closure, electrostatic gyrokinetics
  & \textbf{Online}
  & Closes the Hermite hierarchy in velocity space; $16\times$ velocity-resolution reduction in the non-linear regime; $<2\%$ linear error. \\

\citet{liu_data-driven_2022}
  & PINN
  & Vlasov-Amp\`{e}re simulations
  & Implicit fluid closure, Landau damping
  & Offline
  & Early PINN application to plasma closure; precursor to \citet{qin_data-driven_2023}. \\

\citet{qin_data-driven_2023}
  & PINN, gPINN, gPINNp
  & Vlasov-Amp\`{e}re simulations
  & Implicit fluid closure, Landau damping
  & Offline
  & gPINNp (pressure gradient penalty) most accurate; extrapolates to late times without late-time training data. \\

\citet{wei_data-driven_2023}
  & FNO (Fourier Neural Operator)
  & Kinetic simulations (multiple ICs)
  & Multi-moment fluid closure, Landau damping
  & Offline
  & FNO achieves 10.8\% error vs.\ 66.7\% for MLP; operator learning generalises across initial conditions. \\

\citet{huang_machine-learning_2025}
  & FNO (Fourier Neural Operator)
  & Vlasov solver simulations
  & Heat flux (Hammett-Perkins type), Landau damping
  & \textbf{Online}
  & FNO integrated into fluid solver; captures linear and non-linear Landau damping; outperforms HP closure. \\

\citet{joglekar_machine_2023}
  & Differentiable PDE sim + feedforward NN (160 params)
  & Vlasov-Fokker-Planck simulations (200 runs)
  & Hidden-variable kinetic closure, Landau damping
  & \textbf{Online}
  & End-to-end online training; generalises to $20\times$ larger domain; $14\times$ speedup vs.\ kinetic solver. \\

\midrule
%%% --- Section 3.3: Equation discovery ---
\multicolumn{6}{l}{\textit{Section 3.3: Data-driven equation discovery}}\\
\midrule

\citet{alves_data-driven_2022}
  & SINDy (integral/weak form)
  & PIC simulations (two-stream)
  & Closure hierarchy (heat flux, pressure anisotropy)
  & Offline
  & Pareto front of accuracy vs.\ complexity; recovers known analytic limits at low complexity. \\

\citet{donaghy_search_2023}
  & Sparse regression
  & PIC (magnetic reconnection)
  & Pressure tensor (on- and off-diagonal)
  & Offline
  & On-diagonal $\mathsf{P}$ well-captured; off-diagonal (agyrotropy) most challenging. \\

\citet{ingelsten_data-driven_2024}
  & Sparse regression (WSINDy)
  & PIC (Langmuir waves, two-stream instability)
  & Heat flux, 1D electrostatic plasma
  & Offline
  & Six-term optimal closure; $>$95\% heat flux variance explained; interpretable and robust. \\

\citet{ingelstenDatadrivenMultispeciesHeat2026}
  & Non-linear sparse regression (NLSR) + NN baseline
  & PIC (OSIRIS, two-stream)
  & Multi-species heat flux, 1D electrostatic
  & Offline
  & Generalises the six-term closure to multiple species; rational NLSR models reach NN-equivalent accuracy. \\
\addlinespace

\citet{mcgrae-menge_embedding_2026}
  & Data augmentation + sparse regression + NN
  & PIC (magnetic reconnection)
  & Pressure tensor closures
  & Offline
  & Symmetry augmentation (Lorentz/Galilean) improves accuracy and data efficiency; outperforms CGL and \citet{le_equations_2009}. \\

\citet{cheng_data-driven_2023}
  & mPDE-Net (deep learning)
  & Vlasov-Amp\`{e}re simulations
  & Multi-moment fluid PDEs including closure, Landau damping
  & Offline
  & Learned fluid PDEs reproduce correct Landau damping rate; demonstrates PDE discovery from kinetic data. \\

\bottomrule
\end{longtable}
}

\section{Discussion}

\subsection{Overview of Prior Work}

Work on applying novel machine learning techniques to the plasma closure problem is ongoing, with the majority of the studies outlined here being benchmark and proof-of-concept tests rather than full-fledged applications of the technique. It is clear from the work outlined here that the techniques are viable for test cases and worthy of investigation into more complex problems.

The earliest studies established feasibility by training networks on known \emph{analytic} closures. \citet{ma_machine_2020} showed that several architectures could reproduce the non-linear Hammett-Perkins heat flux \citep{hammett_fluid_1990}, and \citet{maulik_neural_2020} extended this across regimes of differing collisionality, surfacing a link between network connectivity and the spatial locality of the target closure that has since guided architecture choice. Attention then shifted to kinetic training data and to embedding closures within running solvers: \citet{wang_deep_2020} provided the first online test of a neural-network closure inside a fluid simulation, while \citet{laperre_identification_2022} were the first study considered here to train on fully kinetic, two-dimensional PIC data, in doing so exposing the difficulty of capturing the off-diagonal pressure-tensor components, a problem that recurs across both neural-network and equation-discovery approaches throughout this review.

In parallel, equation discovery emerged as an interpretable alternative. \citet{alves_data-driven_2022} brought sparse regression to fully kinetic PIC data, introducing an integral (weak-form) formulation to suppress particle noise and a Pareto hierarchy of reduced models that ranks closure approximations by their contribution to accuracy; \citet{donaghy_search_2023} carried the same approach to a ten-moment reconnection closure, and \citet{liu_data-driven_2022} demonstrated a physics-informed neural network as an inverse-problem alternative for recovering the governing equations. Physics-informed methods also proved valuable for extrapolation: \citet{qin_data-driven_2023} showed that a gradient-enhanced PINN \citep{yu_gradient-enhanced_2022} trained only on early-time data could predict Landau damping at late times, evidence that even a modest use of the underlying physics can extend a model beyond its training window.

More recent studies have expanded the scope of the field considerably. \citet{wei_data-driven_2023} demonstrated the potential of neural operators for plasma closure by applying a Fourier Neural Operator (FNO) to learn from multiple initial conditions simultaneously, while \citet{huang_machine-learning_2025} integrated an FNO closure into a fluid solver, achieving one of the first online demonstrations of a data-driven closure surpassing the Hammett-Perkins model for both linear and non-linear Landau damping. \citet{joglekar_machine_2023} took a fundamentally different approach by training a compact neural network end-to-end through the dynamics of a differentiable fluid solver, demonstrating that online training can yield generalisable closures even with an extremely compact network architecture. On the equation discovery side, \citet{ingelsten_data-driven_2024} extended sparse regression to identify the relative importance of individual terms in the discovered closure, while \citet{mcgrae-menge_embedding_2026} showed that embedding physical symmetries via data augmentation can significantly improve the accuracy and physical consistency of data-driven closures. The study by \citet{miloshevich_electron_2025} advanced neural network closures to the more physically realistic setting of the turbulent magnetosheath, demonstrating that convolutional architectures can generalise across simulation fidelity levels and plasma parameters. A complementary direction was taken by \citet{barbour_machine-learning_2025}, who closed the moment hierarchy in velocity space using a reservoir-computing network embedded online within a Fourier-Hermite kinetic solver.

\subsection{Comparison of Methods}

The progress of work on applying scientific machine learning principles to the plasma closure problem can be broadly divided into two methodological families: neural network surrogates, which learn a closure relation as a numerical mapping, and equation discovery methods, which seek an explicit analytic expression for the closure via sparse or symbolic regression. These two families represent fundamentally different trade-offs in expressivity, interpretability, and computational cost, and the choice between them depends on the goals of the modelling effort.

Neural networks (including standard feedforward architectures, convolutional neural networks, and neural operators such as the FNO and DeepONet) offer strong capacity for function approximation and are able to learn non-local and non-linear closure relations that analytic models struggle to capture. Forward evaluation of a trained network is typically very fast and efficient, making neural network closures attractive for integration into fluid solvers. However, the `black box' nature of neural network models limits physical interpretation, a key consideration when predictive modelling must be consistent with known plasma theory. Standard neural networks also have a tendency to overfit to the training distribution and may not generalise reliably to untested regimes. As noted in Section~2.2.1, network closures extrapolate poorly beyond their training distribution, whereas sparse-regression closures constrained to an interpretable functional form may capture parameter extremes more reliably; this is one of the central trade-offs between the two approaches.

Physics-informed neural networks (PINNs) address some of these shortcomings by constraining the model with conservation laws and symmetries of the system, which can improve performance in sparse data regimes and enable extrapolation beyond the training set \citep{raissi_physics-informed_2019}. However, training PINNs requires computation of higher-order derivatives at each epoch during the automatic differentiation for backpropagation, which can dramatically increase the computational cost of training. PINNs also frequently fail to capture highly varying or oscillatory structure, such as those associated with wave-particle interactions in plasma, due to the resolution required to compute the physics loss terms accurately. A further practical limitation is that PINNs that learn the entire system of fluid equations jointly, as in the work of \citet{qin_data-driven_2023}, produce an implicit closure that cannot be extracted as a standalone module and inserted into an independent fluid solver.

Neural operators take a different approach by learning the mapping between function spaces, which in principle allows the model to capture the underlying physics over a range of initial and boundary conditions without retraining \citep{kovachki_neural_2024}. This property is particularly attractive for closure modelling, where the closure must remain accurate across different plasma states and geometries. Neural operators have also been shown to possess a `discretisation invariance' that allows training on coarser grids and evaluation on finer ones. However, neural operators typically have higher data requirements and can be slower to converge during training compared to standard neural networks. While neural operators reduce dependence on the exact discretisation, accuracy at finer resolutions than those used in training is not guaranteed, placing practical limits on their ability to extrapolate beyond training conditions. A key practical advantage of the FNO architecture is that the FFT-based kernel maps efficiently to GPU hardware, making the evaluation of non-local closures competitive with local analytic closures in terms of computational cost addressing one of the longstanding objections to Hammett-Perkins-type closures in large-scale codes.

Equation discovery methods produce interpretable, closed-form expressions that allow direct physical insight, making them valuable for hypothesis generation and model reduction. Because the discovered closure takes the form of an explicit mathematical expression, it can be analysed for physical consistency, for example, by checking whether it satisfies expected symmetries of the underlying equations or generates positive entropy production. As reviewed in Section~3.3.1, such symmetry constraints can be enforced during training (e.g.\ via the data augmentation of \citet{mcgrae-menge_embedding_2026}), a guarantee that equation-discovery closures accommodate naturally but network surrogates do not.

As discussed in Section~2.2.2, equation discovery encompasses two broad strategies with distinct trade-offs. Sparse regression, as implemented in the SINDy framework \citep{brunton_discovering_2016} and its variants, selects the most relevant terms from a predefined library of candidate functions and is computationally efficient when a suitable library can be constructed from prior physical knowledge. Its main drawback is that the library must be specified in advance, and the method is limited to discovering expressions within the span of the chosen library; if the true functional form lies outside this space, the model will be inaccurate. General symbolic regression, by contrast, searches over a much larger space of mathematical expressions using stochastic strategies such as genetic programming \citep{cranmer_discovering_2020}, without requiring a fixed library. This greater flexibility comes at a higher computational cost, but allows the discovery of functional forms that would not be found by sparse regression. These cross-domain successes, and the recent LLM-hybrid approach of \citet{ying_neural_2025} are discussed in section~2.2.2.

To date, the plasma closure studies reviewed here have relied exclusively on sparse regression; general symbolic regression has not yet been applied to the plasma closure problem. This represents an important gap in the literature. Symbolic regression could be particularly valuable for discovering closure relations in regimes where prior physical knowledge is insufficient to construct an adequate candidate library, for example, in strongly turbulent or multi-species plasmas where the functional form of the closure is largely unknown. The development of GPU-accelerated symbolic regression tools and the integration of large language models into the search process \citep{ying_neural_2025} may make this approach increasingly tractable for the high-dimensional problems encountered in plasma physics.

An emerging strategy is to combine neural networks and equation discovery in a complementary workflow. As demonstrated by \citet{ingelstenDatadrivenMultispeciesHeat2026} and briefly discussed in Section~\ref{sec:data_eq_disco}, neural networks can serve as a useful preliminary step in the equation discovery process: by first training a neural network to confirm that a sufficiently accurate model exists within a given input space, the neural network provides an accuracy baseline against which the fidelity of the sparser regression model can be judged. This two-stage approach is a promising direction for future closure development. In particular, symbolic regression could be deployed in the second stage when the library-based sparse regression approach fails to capture the correct functional form, leveraging the neural network benchmark to guide the search through a broader expression space.

Overall, these methods can be viewed as complementary rather than competing strategies. Neural networks and neural operators excel in expressivity and predictive performance, making them well suited for data-rich, strongly non-linear regimes. Physics-informed networks strike a balance between data-driven flexibility and physical consistency, while sparse regression provides transparent models that can be related to the underlying physics of the system. For collisionless plasma modelling, an optimal closure strategy may involve hybrid approaches that blend these methodologies, thereby capturing the interpretability of analytic closures while retaining the expressivity of machine learning models.

\subsection{Ongoing Challenges}

The goal of improving the plasma closure model is to attain a data-driven closure model that can be integrated into fluid simulations to include kinetic physics in large-scale plasma fluid simulations, leading to improved predictions and forecasting ability.

The challenges currently faced in learning closure models from kinetic data are, in essence, the challenges of most machine learning applications. In addition to the common difficulties of machine learning approaches, there are the challenges presented by the requirement for our data-driven closure models to be integrated into large-scale plasma models. The learned model must be numerically stable within the large-scale plasma fluid and be able to predict forward in time without significant growth of numerical error. The issue of including multiple different initial conditions must also be addressed by future research; for the learned model to be useful when implemented into a large-scale fluid model, it must be able to adapt and maintain accuracy for the different phenomena and conditions that would be present in a large-scale three-dimensional plasma fluid.

\subsubsection{Limitations to specific parameter ranges}

A fundamental limitation of all data-driven closure models, whether neural network surrogates or discovered analytic expressions, is that they are trained on data drawn from a finite region of parameter space and cannot be expected to remain accurate outside this region. Neural networks, in particular, are well-known to be unreliable when extrapolating beyond the training distribution: a network trained on data from a specific range of plasma beta, collisionality, or perturbation amplitude may produce physically meaningless results when evaluated at parameter values not represented in the training set. This limitation has practical consequences for any attempt to deploy a data-driven closure in a large-scale simulation where the plasma conditions may vary widely across the domain. Including a greater range of parameters within training data would require greater amounts of training data (which could be expensive if sourced from kinetic simulations) and would require a greater numerical complexity of the model.

Even within the training distribution, accuracy is not guaranteed uniformly. The studies reviewed here have shown that closure models tend to perform best in regimes well-represented in the training data and degrade in under-sampled regions, such as the off-diagonal components of the pressure tensor in reconnection geometries \citep{laperre_identification_2022, donaghy_search_2023}. Sparse regression closures face an analogous limitation: because the discovered expression is drawn from a predefined library of candidate terms, the model can only represent physics that lies within the span of the library, and any regime requiring terms outside this space will be inaccurately described.

Addressing this limitation will require both methodological advances, such as physics-informed constraints that enforce known conservation laws and symmetries beyond the training distribution, and practical strategies for characterising the domain of validity of a trained closure. Uncertainty quantification techniques, which can highlight predictions in under-represented parameter regions, could be valuable in this context but have not yet been explored in the plasma closure literature.

\subsubsection{Accurate models}

The investigations conducted to date have been able to report high levels of accuracy. Previous work has been chiefly concerned with the development of proof-of-concept models that have consistently achieved higher accuracy than the baseline and analytic closure models they were compared to. However, previous studies rarely account for any higher-dimensional cases, where, for the few investigations that did attempt higher-dimensional cases (such as \citet{laperre_identification_2022} and \citet{donaghy_search_2023}), difficulties in modelling the off-diagonal components of the pressure and heat flux tensors were found. Future work, therefore, must be concerned with the improvement of modelling accuracy for these off-diagonal components, for higher-dimensional cases, as well as more complex dynamics, with the majority of research thus far being concerned with simple one-dimensional linear Landau damping.

An important and largely unexplored direction for future work is the systematic assessment of the physical properties of discovered closure models. For a data-driven closure to be trusted in a simulation, it must not only fit the training data well but also satisfy fundamental physical constraints: conservation of energy and momentum, Galilean (or Lorentz) invariance, and positive-definite entropy production. These properties are rarely checked in the current literature. The data-augmentation approach of \citet{mcgrae-menge_embedding_2026} (Section~3.3.1) is a promising step in this direction. More broadly, future work should investigate whether discovered closures preserve the well-posedness and stability of the fluid system into which they are embedded, for instance, whether the augmented fluid equations remain hyperbolic and whether the closure introduces spurious growing modes.

\subsubsection{Generalisable models}

Some investigation has been conducted into the ability of physics-informed neural networks to allow for extrapolation beyond the training data used. Work from \citet{qin_data-driven_2023} was able to show that a PINN trained on only simulation data sampled from early times was able to accurately predict Landau damping at late times. There are, however, still difficulties in achieving this kind of extrapolation for more complex dynamics, especially for high-frequency behaviour that the PINN would struggle to model. The example from \citet{qin_data-driven_2023} was also only for a one-dimensional problem; there has been no investigation into the accuracy of models beyond their training data for higher-dimensional problems. Going forward, it will be important to keep exploring the ability of physics-informed machine learning to supplement training data to allow for the models to accurately predict behaviour outside of the region covered by the training data, especially for more complex phenomena.

\subsubsection{Inclusion of multiple different phenomena and conditions}

For the implementation of data-driven models into large-scale simulations, a model will need to be able to respond to varying conditions at its boundaries and to different plasma regimes. The majority of research has utilised neural networks that require re-training for the inclusion of different initial or boundary conditions. Early work on applying neural operators, which could potentially handle different conditions without retraining, has been undertaken by \citet{wei_data-driven_2023}. The differentiable-simulation approach of \citet{joglekar_machine_2023}, which trains the closure end-to-end through the dynamics of the fluid solver, offers another route: because the training loss is evaluated on full simulation trajectories rather than individual snapshots, the resulting closure is implicitly optimised for stability and accuracy over extended time evolution. Future work should investigate whether neural operators and differentiable-simulation approaches can achieve good accuracy across different initial and boundary conditions for more complex phenomena found in higher-dimensional cases; these are not the only options, however, and investigation into other alternatives (such as deploying multiple specialised closures selected according to local plasma conditions) will be required for comparison.

\subsubsection{Implementation into large-scale plasma fluid simulations}

The work from \citet{wang_deep_2020} was one of the first to implement a data-driven closure model into a fluid simulation; it was found that the numerical error did not grow significantly, and the data-driven model did perform better at late times than the traditional numerical model. The work from \citet{huang_machine-learning_2025} subsequently integrated an FNO closure model of Landau damping into a fluid solver, showing the data-driven closure to improve on the Hammett-Perkins result for both linear and non-linear Landau damping. \citet{joglekar_machine_2023} demonstrated a further advance by training the closure model directly through the fluid solver, achieving stable online performance and generalisation to domains much larger than the training set. All three addressed only one-dimensional Landau damping. Future work will need to investigate the ability for these data-driven models to be integrated into plasma fluid simulations in higher dimensions and larger-scale systems to discover how numerically stable they can remain, and to give an estimate of the time frames over which these models could be implemented while retaining the correct kinetic physics.

\subsubsection{Kinetic simulation data for training}

The different studies considered thus far have generated their kinetic simulation data, whether from a PIC code, Vlasov solver, or other. Generating new simulation data, while useful for obtaining simulation data for the specific case being investigated, does require large computational resources and for future research into two- and three-dimensional cases, as well as for the possible inclusion of multiple different phenomena and conditions, this will represent a steep computational cost. One possibility to combat the computational cost would be to aggregate kinetic simulation data from multiple sources, for example, prior simulation data from different research groups, as opposed to generating all the data individually, to leverage the community to obtain a larger volume of training data, therefore improving future models. It will also be important in future research to include observational data from satellite missions in the training of the closure models. However, this will present many challenges; the exact phenomenon will not be as clearly defined as is possible in simulation.

\section{Acknowledgments}

This work has been supported by a Principal's award studentship from Queen Mary University of London, and by NASA grant 80NSSC26K0086. The authors declare no competing interests.

\appendix
\section{Machine Learning Glossary}

Provided here is a brief glossary of machine learning terms used throughout this review paper for plasma physicists who may be unfamiliar with the field.

\subsection{General Concepts}

\begin{itemize}
    \item \textbf{Machine Learning (ML):} A subset of artificial intelligence (AI) focused on building models that learn from data to make predictions or decisions without being explicitly programmed.
    \item \textbf{Data-Driven Approach:} Using data (often from simulations or observations) to inform or construct models, rather than relying solely on theoretical or analytic formulations.
    \item \textbf{Model:} A mathematical or computational representation trained to map inputs to outputs, used for prediction or analysis.
    \item \textbf{Dataset:} A structured collection of data used for training, validating, or testing ML models.
    \item \textbf{Deep Learning:} The process of training neural network models with more than one hidden layer. It can also refer to the general field of neural networks with multiple hidden layers.
\end{itemize}

\subsection{Types of ML}

\begin{itemize}
    \item \textbf{Supervised Learning:} Training ML models on data with paired sets of inputs (features) and outputs (labels), typically used for regression and classification tasks.
    \item \textbf{Unsupervised Learning:} Models are trained on unlabelled data, aiming to find underlying patterns (e.g., clustering, dimensionality reduction).
    \item \textbf{Semi-Supervised Learning:} Combines labelled and unlabelled data for training, often used when labelled data is scarce.
    \item \textbf{Reinforcement Learning:} A type of ML where an agent learns to make decisions by receiving rewards or penalties for its actions.
\end{itemize}

\subsection{Neural Network Concepts and Architectures}

\begin{itemize}
    \item \textbf{Neural Network:} A computational model inspired by the human brain, composed of layers of interconnected nodes (neurons) capable of learning complex patterns.
    \item \textbf{Input Layer:} The first layer of a neural network, which receives the feature vector.
    \item \textbf{Hidden Layer:} The layers between the input and output layers of a neural network.
    \item \textbf{Output Layer:} The final layer of a neural network, producing the model's predictions.
    \item \textbf{Multi-Layer Perceptron (MLP):} A type of feedforward neural network with multiple layers of neurons, typically used for regression or classification.
    \item \textbf{Convolutional Neural Network (CNN):} A neural network architecture particularly suited for processing data with grid-like topology (e.g., images, multidimensional simulation data).
    \item \textbf{Physics-Informed Neural Network (PINN):} A neural network that incorporates physical laws (e.g., differential equations) into its training process, often by including them in the loss function \citep{raissi_physics-informed_2019}.
    \item \textbf{Gradient-Enhanced PINN (gPINN):} A variation of PINN that includes gradients of the physics loss terms as additional constraints during training \citep{yu_gradient-enhanced_2022}.
    \item \textbf{Neural Operator:} A neural network designed to approximate mappings between function spaces (operators), rather than just functions. Useful for learning solutions to partial differential equations across varying conditions \citep{kovachki_neural_2024}.
    \item \textbf{Fourier Neural Operator (FNO):} A neural operator that leverages Fourier transforms to learn mappings in problems with periodicity or spatial structure efficiently \citep{li_fourier_2021}.
    \item \textbf{Deep Operator Network (DeepONet):} A neural operator architecture designed to learn non-linear operators for differential equations \citep{lu_deeponet_2020}.
    \item \textbf{Physics-Informed Neural Operator (PINO):} Combines neural operators with physics-informed constraints for learning PDE solutions \citep{li_physics-informed_2023}.
\end{itemize}

\subsection{ML Techniques}

\begin{itemize}
    \item \textbf{Regression:} Predicting a continuous output variable from input data. In the document, this often refers to learning closure relations.
    \item \textbf{Classification:} Assigning input data to discrete categories or classes.
    \item \textbf{Sparse Regression:} A regression technique that selects a minimal set of terms from a large library of candidate functions to best fit the data, promoting interpretability and parsimony.
    \item \textbf{Symbolic Regression:} The process of finding mathematical expressions that best fit a dataset, often via evolutionary algorithms or sparse regression.
    \item \textbf{Pareto Optimisation:} A principle where models are selected based on a trade-off between complexity and accuracy, seeking the simplest model that achieves acceptable error.
    \item \textbf{Equation Discovery:} Using machine learning (often sparse regression or PINNs) to infer governing equations from data.
    \item \textbf{Surrogate Model:} A simplified model (often ML-based) that approximates a more complex or computationally expensive model, used to speed up simulations or enable real-time predictions.
\end{itemize}

\subsection{Training and Evaluation}

\begin{itemize}
    \item \textbf{Training Data:} The portion of a dataset used to fit model parameters.
    \item \textbf{Validation Data:} Used during training to tune model hyperparameters and prevent overfitting.
    \item \textbf{Test Data:} Data not seen during training, used to evaluate model performance and generalisation.
    \item \textbf{Label:} The target output variable in supervised learning.
    \item \textbf{Feature:} An individual measurable property or characteristic used as input to a model.
    \item \textbf{Feature Scaling:} Adjusting the range of input variables to improve model training and convergence.
    \item \textbf{Loss Function:} A mathematical function quantifying the difference between predicted and true values, used to guide model training.
    \item \textbf{Model Parameter:} The numerical values that make up the computations of the model. These are iteratively altered during training to fit the model to the training data. For neural networks, the parameters are the respective weights and biases of each neuron.
    \item \textbf{Early Stopping:} Stopping training when performance on the validation set stops improving to avoid overfitting.
    \item \textbf{Batch Size:} The number of training examples used in one iteration of model parameter updates.
    \item \textbf{Generalisation:} The ability of a model to perform well on data unseen during training, not just the training set.
    \item \textbf{Overfitting:} When a model is too complex and learns the training data too well, including any present noise, and performs poorly on new data, leading to poor generalisability.
    \item \textbf{Underfitting:} When a model is too simple to capture the underlying patterns in the data, resulting in poor performance on both the training and test data.
\end{itemize}

\bibliographystyle{jpp}

\bibliography{references}

\end{document}